\documentclass[aps,pra,twocolumn,superscriptaddress,amsmath,showpacs,amssymb]{revtex4} 

\usepackage{graphicx}
\usepackage{amsthm,amsmath}

\newcommand{\bm}[1]{\mbox{\boldmath{$#1$}}}
\newcommand{\be}{\begin{eqnarray}}
\newcommand{\ee}{\end{eqnarray}}

\newcommand{\nn}{\nonumber}

\newcommand{\bo}{\boldsymbol}
\newcommand{\lb}{\label}

\newcounter{ichi}
\newcounter{ni}
\newcounter{san}
\newcounter{yon}
\newcounter{go}
\newcounter{roku}
\newcounter{nana}
\newcounter{hati}
\newcounter{kyu}
\begin{document}

\preprint{APS/123-QED}

\title{Quantum Field Theoretical Analysis on Unstable Behavior of Bose-Einstein Condensates in Optical Lattices}

\author{K.~Kobayashi}
\email{keita-x@fuji.waseda.jp}
\affiliation{Department of Materials Science and Engineering,
Waseda University, Tokyo 169-8555, Japan}
\author{M.~Mine}
\email{mine@aoni.waseda.jp}
\affiliation{Department of Physics, Waseda University,
 Tokyo 169-8555, Japan}
\author{M.~Okumura}
\email{okumura.masahiko@jaea.go.jp}
\affiliation{CCSE, Japan Atomic Energy Agency, 6-9-3 Higashi-Ueno, Taito-ku, Tokyo 110-0015, Japan
}
\affiliation{CREST(JST), 4-1-8 Honcho, Kawaguti-shi, Saitama 332-0012,
Japan} 
\author{Y.~Yamanaka}
\email{yamanaka@waseda.jp}
\affiliation{Department of Electronic and Photonic Systems,
Waseda University, Tokyo 169-8555, Japan}

\date{\today}

\begin{abstract}   
We study the dynamics of Bose-Einstein condensates flowing
 in optical lattices on the basis of quantum field theory. 
For such a system, a Bose-Einstein condensate shows a unstable behavior
 which is called the dynamical instability. The unstable system is
 characterized by the appearance of modes with complex eigenvalues.
Expanding the field operator in terms of excitation modes including
 complex ones, we attempt to diagonalize the unperturbative Hamiltonian
 and to find its eigenstates.  It turns out that although 
the unperturbed Hamiltonian  is not diagonalizable in the conventional
 bosonic representation the appropriate choice of physical states leads to 
a consistent formulation. Then we analyze the dynamics of the system
in the regime  of the linear response theory.
Its numerical results are consitent with as those given by
the discrete nonlinear Schr\"odinger equation.
\end{abstract}

\pacs{03.75.Lm,03.70.Kk,11.10.-z}
\maketitle

\section{\label{sec:level1}Introduction}
The Bose-Einstein condensates (BECs) of trapped atoms, first realized in
1995 \cite{Cornell,Cornell2,Ketterle2}, are ideal systems for studying
quantum many-body phenomena. This is because the systems are dilute,
weakly interacting ones and we can easily control the configuration of 
the trap and even the strength of atomic interaction. Recently much attention has
been focused on the BEC in an optical lattice, which provides
us with rich phenomena such as superfluid-insulator quantum phase
transition \cite{Greiner} and Bloch oscillations \cite{Bloch}, and 
also enables us to observe directly the quantum fluctuation as
quantum depletion from the condensate \cite{depletion}.

Fluctuations and excitation spectra of a BEC are determined from the
Bogoliubov-de Gennes (BdG) equation, which is obtained by the linearization of the time dependent Gross-Pitaevskii (TDGP) equation. Numerical calculation
of the excitation spectra of a BEC in an optical lattice was performed 
in Ref.~\cite{Ichioka}. The analytical studies under the tight-binding
approximation \cite{Oosten,Clark} gave the theoretical result in a good 
agreement with the experiment \cite{depletion}.

In this paper, we focus on the BECs flowing in optical lattices. 
For such a system, two types of instability are known: Landau and dynamical
 instabilities. The Landau instability is an energetic instability which
is caused by negative energies of the quasi-particle \cite{Iigaya}.
 The dynamical
instability is that of condensates caused by the complex eigenvalues 
of the BdG equations \cite{Wu1,Wu2}. The Landau instability (energetic
instability) is not observed at low temperature \cite{instaove2}, because
it requires the thermal cloud which brings a dissipative mechanism and
drives the condensate toward a lower energy state. On the other hand, the
dynamical instability can occur even at zero temperature. 

The emergence of complex eigenvalues in the BdG equations is not restricted to
the superfluidity of BECs in optical lattices, but is found widely, such as
in the BECs with highly quantized 
vortices \cite{Pu,Garay,Skryabin,Mettenen,Kawaguchi} or gap 
solitons \cite{Hilligsoe}, or in the multi-component BECs \cite{Zhang,Robert}.
In all the cases, a naive scenario of the dynamical instability is as follow:
When the complex eigenvalues appear, the small deviations from the static
condensation have the time-dependence with the complex frequency and
will grow exponentially, and consequently the initial configuration of the
static condensate is destroyed. The dynamical instability of the superfluidity
flowing in an optical lattice was observed at the experiments and the decay
rates of the condensate are well reproduced by the theory based on the TDGP
equation \cite{instaove2,instaove}.  In the case of a highly quantized
 vortex of
BECs, the imaginary part of the complex eigenvalue is considered to be related
 to the experimentally observed lifetime of the vortex \cite{Kawaguchi}.
It is also mentioned that the numerical result of the TDGP is in a
 good agreement with the experimental data \cite{Mateo}. 

The c-number theories seem to describe
the dynamical instability well. But there
are quantum fluctuations even at zero-temperature. So a consistent
treatment of the BEC in full quantum field theory (QFT) is needed for
describing the unstable behavior with quantum fluctuations.
Because of the presence of trapping potentials, this QFT becomes one for finite
volume systems in which the spatially translational invariance is lost and
all the energy levels are discrete. 
The Bose-Einstein condensation is considered as a spontaneous breakdown of
a global phase symmetry, so the Nambu-Goldstone (NG) mode inevitably
appears \cite{Nambu}.
This zero-mode requires us subtle treatments
\cite{Lewenstein,Matsumoto,Okumura1}, closely related to its observable effects
\cite{Okumura2,MKOY} and several theoretical matters such as the inequivalent
vacua, Ward-Takahashi relations and Hugenholtz-Pines theorem
 \cite{Okumura3,MOY,Okumura4,Enomoto}.

It is desirable to include complex modes into the QFT in addition to 
the zero-mode.
We have attempted to formulate QFT in the presence of complex-energy
modes  in the case of highly quantized vortices in the previous
 paper \cite{MOSY}.
In this paper we present a consistent formulation of QFT with complex modes
for the BECs flowing in an optical lattice . These formulations in full quantum theory are expected to go beyond the conventional c-number approach 
of the mean-field approximation and the BdG equations.

Needless to say, QFT in nonequilibrium  situations is needed in wide area, 
for example, in particle physics, cosmology and condensed matter physics, 
and its construction is still a challenging subject.
The QFT description of unstable behaviors, presented here, may 
become important for some nonequilibrium systems.

This paper is organized as follows.
 In Sec.~II, the model action and Hamiltonian are given.
 We assume that the solution of the Gross-Pitaevskii (GP) equation satisfies
 the Bloch condition. 
 In Sec.~III, we expand the field operator 
 in terms of the adequate complete orthonormal set
 and make the tight-binding approximation in the
 unperturbed Hamiltonian.
 In Sec.~IV, solving the equation of time evolution  for the 
operators, we find eigenmodes, some of whose eigenvalues may be complex.
We show that the operators in the complex mode sector are not
subject to the usual bosonic commutation relations and that
the unperturbed Hamiltonian can not be diagonalized
 in the usual bosonic representation.
One can check that the canonical commutation relations
for the field operators which are the fundamental requirements in QFT
are kept  under the tight-binding approximation
and in the presence of complex modes.
 Next, we obtain the eigenstates of complex modes and their properties.
In Appendix A, we summarize
the eigenfunctions of the Bogoliubov-de Gennes (BdG) equation,
in connection with the discussion in Sec.~IV.
 In Sec.~V, we introduce the conditions for physical states
 which provide us with a consistent QFT
 description of the unstable behavior.
 In Sec.~VI, applying Kubo's linear response theory (LRT)
 to the physical states introduced in Sec~V,
 we calculate the density
 response of the system against the external perturbation,
whose lengthy expression is given in Appendix B.
 Our numerical results of the density
 response reproduce those of  the discrete nonlinear
 Schr\"odinger equation (DNSE), which is obtained by applying
the tight-binding approximation to the TDGP equation \cite{DNLS,DNLS2}.
 Section~VII is devoted to summary.

\section{\label{sec:level2}Model Hamiltonian of Quantum Field Theory for BEC in Optical Lattice}
We consider the trapped BEC of neutral atoms in an optical lattice.
In QFT, the model action which describes the system is given by
\be
S&=&\int \! dt \, d^{3}x
\Big{
\{}\Psi^{\dagger}(x)\big{(}T-K-V_{\rm{opt}}+\mu\big{)}\Psi(x) \nonumber \\
& &\qquad\qquad-\frac{g}{2}\Psi^{\dagger}(x)\Psi^{\dagger}(x)\Psi(x)\Psi(x)\Big{\}} \, ,
\label{action}
\ee
with $x=(\bm{x},t)$.
Here we use the following notations:
\be
T&=&i\hbar\frac{\partial}{\partial t}  \\
K&=&-\frac{\hbar^{2}}{2m}\nabla^{2}  \\
V_{\rm{opt}}(\bo{x})&=&
\sum_{i=x,y,z}V_{0}\cos^{2} \left(\frac{2\pi}{d_{i}} i \right)
\, . 
\ee 
The parameters  $\mu$, $m$ and $g$ represent the chemical potential, the mass of the neutral atom
and the coupling constant of atomic interaction, respectively. 
The optical
potential whose strength is denoted by $V_{0}$ has the lattice spacing
 $d_{i}$ ($i=x,y,z$) in each direction, 
that is, $V_{\rm{opt}}(\bo{x+G})=V_{\rm{opt}}(\bo{x})$, where $\bo{G}$ is a 
Bravais lattice vector, $\bo{G}=(l_{x}d_{x},l_{y}d_{y},l_{z}d_{z})$ 
with integers $l_{i}$ ($i=x,y,z$). 
We consider the situation without the harmonic potential for atoms. 

The action (\ref{action}) is invariant 
under the global phase transformation  $\Psi(x) \to e^{i\theta}\Psi(x) $ 
and  $\Psi^{\dagger}(x)  \to e^{-i\theta}\Psi^{\dagger}(x)$,
where $\theta$ is an arbitrary real constant. When a BEC is created,
the global phase symmetry is spontaneously broken.  In the terminology of the operator formalism
(canonical formalism) for quantum field theory (QFT), the Heisenberg field
$\hat{\Psi}(x)$ is then divided into a c-number part $v(\bo{x})$ 
and an operator one $\hat{\phi}(x)$,
\be
{\hat \Psi}(x)= v({\bm x})+{\hat \phi}(x) \, .
\ee
The c-number field  $v(\bo{x})$, called the order parameter,
is assumed to be time-independent throughout this paper.
The order parameter is defined
as an expectation value of the  Heisenberg field with respect to the vacuum $|\Omega\rangle$,
\be
\langle\Omega|\hat{\Psi}(x)|\Omega\rangle=v(\bo{x}) ,
\ee
or equivalently
\be
\langle\Omega|\hat{\phi}(x)|\Omega\rangle=0 \lb{eq;phi}.
\ee

Let us introduce an additional symmetry breaking term \cite{Okumura1}
\be
\triangle S=\int \! dt \, d^{3}x \varepsilon\bar{\epsilon}\left[v^{*}(\bo{x})\Psi(x)+v(\bo{x})\Psi^{\dagger}(x)\right] \, ,
\lb{eq:deltaS}
\ee
and the total action $S_{\varepsilon}$ is defined by
\be
S_{\varepsilon}=S+\triangle S \, .
\ee
Here $\varepsilon$ is an infinitesimal dimensionless parameter
 and $\bar{\epsilon}$ represents a typical energy scale of the system.
The total action $S_{\varepsilon}$ is not invariant under the global phase transformation of $\Psi(x)$, due to the additional breaking term.

The above method corresponds to the Bogoliubov's quasi-average,
known in the treatment of spontaneously broken systems \cite{BogoliubovL}.
The breaking term specifies the ``direction'' of the symmetry breakdown,
so a corresponding vacuum is selected among many degenerate ones. 
The singularity associated with the Nambu-Goldstone (NG) mode \cite{Nambu}
is regularized \cite{Okumura1}.
At the final stage of calculation, the limit of $\varepsilon \rightarrow
0$ is taken, and original symmetry is restored.

We move to the interaction representation of the canonical formalism. 
The canonical commutation relations (CCRs) for the field operators
are given as follows:
\be
\big{[}\hat{\phi}(\bo{x},t),\hat{\phi}^{\dagger}(\bo{x'},t)\big{]}=\delta(\bo{x-x'})
\, ,\\
\big{[}\hat{\phi}(\bo{x},t),\hat{\phi}(\bo{x'},t)\big{]}=\big{[}\hat{\phi}^{\dagger}(\bo{x},t),\hat{\phi}^{\dagger}(\bo{x'},t)\big{]}=0 \, .
\ee 
The total Hamiltonian of the system is divided into the two terms,
\be
\hat{H}=\hat{H}_{0}+\hat{H}_{\mathrm{int}} \, ,
\ee
where the unperturbative Hamiltonian $\hat{H}_{0}$ and the interaction one
$\hat{H}_{\mathrm{int}}$ are given as
\be
\hat{H}_{0} &=& \int \! d^{3}x \Big{[} \hat{\phi}^{\dag} (K+V-\mu)
\hat{\phi} \nonumber \\ 
& &{}+\frac{g}{2}\big{(} 4 |v|^{2} \hat{\phi}^{\dag} \hat{\phi} + v^{*2}
\hat{\phi}^{2}+v^{2}\hat{\phi}^{\dag 2}\big{)} \Big{]} \lb{eq;hami} \, ,
\\ 
\hat{H}_{\mathrm{int}} &=& \int \! d^{3}x \Big{[} v^{*}
\left(K+V-\mu+g|v|^{2} - \varepsilon \bar{\epsilon} \right) \hat{\phi}
\nonumber \\ 
& & {} + \hat{\phi}^{\dag} \left(K+V-\mu+g|v|^{2} - \varepsilon
\bar{\epsilon} \right) v \nonumber \\
& & {} + g \left( v\hat{\phi}^{\dag 2} \hat{\phi} + v^{*}
\hat{\phi}^{\dag} \hat{\phi}^{2} \right) + \frac{g}{2} \hat{\phi}^{\dag
2} \hat{\phi}^{2} \Big{]}  \, , 
\end{eqnarray}
respectively.  The renormalization counter terms are suppressed.

The condition (\ref{eq;phi}) at the tree level (zero-loop level) derives the following
classical equation for $v(\bo{x})$:
\be
\left(K+V_{{\rm opt}}+g|v(\bo{x})|^{2}-\mu-\varepsilon\bar{\epsilon}\right)v(\bo{x})=0 \lb{eq;GP}\, . 
\ee
This equation is nothing but the Gross-Pitaevskii (GP)
 equation \cite{GP} at the limit of
$\varepsilon \to 0$.  The quantity $|v(\bo{x})|^{2}$ is identified with
the density of condensed particles $n(\bo{x})$ as $n(\bo{x})=|v(\bo{x})|^{2}$.
The total condensate particle number $N_{\rm c}$ is given by
$N_{\rm c}=\int \! d^{3} \, x|v(\bo{x})|^{2}$.

\subsection{\label{sec:level3}Bloch condition for Gross-Pitaevskii equation}
In this paper we assume that the density of condensate particles
$n(\bo{x})=|v(\bo{x})|^{2}$ has the same periodicity as the lattice as
$n(\bo{x+G})=n(\bo{x})$, although it may not be periodic strictly.
Under the assumption of this periodicity, the solution of the GP equation (\ref{eq;GP})
is expressed by a function satisfying the Bloch condition \cite{taylor}: 
\be
v_{\bo{k}}(\bo{x})&=&e^{i\bo{k\cdot x}}\bar{v}_{\bo{k}}(\bo{x})\qquad
\left( k_{i}=\frac{2\pi l_{i}}{L_{i}} \right) \, , \\
\bar{v}_{\bo{k}}(\bo{x+G})&=&\bar{v}_{\bo{k}}(\bo{x})\, .
\ee 
Hereafter we specify the solution of the GP equation by the vector $\bo{k}$ as $v_{\bo{k}}(\bo{x})$.
Here $l_{i}$ ($i=x,y,z$) ia an integer and $L_i$ represents
a length of the system in each direction. 

In this paper, we are interested in the solution of the GP equation with
the condensate flow in an optical lattice.
Only when $v_{\bo{k}}(\bo{x})$ is complex, the condensate can have a flow.
The velocity of the condensate flow is given by
\be
v_{\bo{k}}=\frac{\hbar}{m} \bo{k} +\frac{\hbar}{m}\bo{\nabla}S_{\bo{k}}(\bo{x})\, ,
\ee
where $S_{\bo{k}}(\bo{x})$ is the phase of $\bar{v}_{\bo{k}}(\bo{x})$.

\section{Tight-binding approximation in optical lattice}
Here we discuss BECs in a cubic lattice for simplicity ($d=d_x=d_y=d_z$).
Consider the following eigenequation:
\be
[K+V_{\rm{opt}}+g|v_{\bo{k}}|^{2}-\mu-\varepsilon\bar{\epsilon}]f_{\bo{kq}}^{(n)}(\bo{x})=\epsilon_{\bo{kq}}^{(n)} f_{\bo{kq}}^{(n)}(\bo{x})\,  .\nonumber\\ \lb{eq:GPeigen}
\ee
The density of the condensate particles
$n_{\bo{k}}(\bo{x})=|v_{\bo{k}}(\bo{x})|^{2}$ has the periodicity of the lattice spacing $d$,
so that one may require $f_{\bo{kq}}^{(n)}(\bo{x})$ to satisfy the Bloch condition as
\be
f_{\bo{kq}}^{(n)}(\bo{x})&=&e^{i\bo{q\cdot x}}\bar{f}_{\bo{kq}}^{(n)}(\bo{x}) \, ,\\
\bar{f}_{\bo{kq}}^{(n)}(\bo{x+G})&=&\bar{f}_{\bo{kq}}^{(n)}(\bo{x})\, ,
\ee
where $\bo{q}$ and $n$ are a Bloch wave vector and  a band index, respectively.
Equation (\ref{eq:GPeigen}) has the zero-energy solution which
is proportional to $v_{\bo{k}}(\bo{x})$ as
\be
v_{\bo{k}}(\bo{x})=\sqrt{N_{\rm c}}e^{i\bo{k\cdot
x}}\bar{f}_{\bo{kk}}^{(1)}(\bo{x}) \lb{zero mode}\, , 
\ee
with $\epsilon^{(1)}_{\bo{kk}}=0$. 
The completeness and orthonormal conditions are
\be
\int \! d^3 x \, f_{\bo{kq'}}^{(n')*} (\bo{x})f_{\bo{kq}}^{(n)}(\bo{x}) 
&=&\delta_{n'n}\delta_{\bo{q'q}}  \lb{eq:Bo}\, , \\
\sum_{n,\bo{q}} f_{\bo{kq}}^{(n)}(\bo{x})f_{\bo{kq}}^{(n)*}(\bo{x'})
&=&\delta(\bo{x}-\bo{x'}) \lb{eq:Bc}\, .
\ee

We consider the situation in which the strength 
of the lattice potential $V_{0}$ is so large that
the wave packet of the condensate wavefunction is localized
in each site of the lattice. The wave packet localized
in each site is represented by a Wannier function, whose set
$\{w_{\bo{ki}}^{(n)}(\bo{x}) \} $ is related to the set
$\{f_{\bo{kq}}^{(n)}(\bo{x})\}$ by the unitary transformation, 
\be
w_{\bo{ki}}^{(n)}(\bo{x})=I_{\rm s}^{-\frac{1}{2}}
\sum_{\bo{q}}e^{-i\bo{q\cdot x_{i}}} f_{\bo{kq}}^{(n)}(\bo{x}) \lb{def
of Wannier}\, , 
\ee
where $I_{\rm s}$ is the total number of lattice sites  and $\bo{x_i}$ is
 the position of the $\bo{i}$-th site,
 where $\bo{i}=(i_x, i_y, i_z)$ with integers $i_j$ $(j=x, y, z)$. 
The Bloch wave vector $\bo{q}$ runs over
the Brillouin zone. 
The completeness and orthonormal condition of the Wannier functions
 is guaranteed by Eqs.~($\ref{eq:Bo}$) and ($\ref{eq:Bc}$),
\be
\int\! d^{3}x\, w_{\bo{ki}}^{(n')*}(\bo{x})w_{\bo{kj}}^{(n)}(\bo{x})&=&\delta_{n'n}\delta_{\bo{ij}}
\label{eq:WaniOrthoNorm}
\, ,\\
\sum_{n,\bo{i}}w_{\bo{ki}}^{(n)*}(\bo{x'})w_{\bo{ki}}^{(n)}(\bo{x})&=&\delta(\bo{x}-\bo{x'}) \, .
\ee
If the density of condensate particle $n_{\bo{k}}(\bo{x})$ 
has the reflection symmetry ($n_{\bo{k}}(-\bo{x})=n_{\bo{k}}(\bo{x})$), 
we can easily derive the following relation 
\be
f_{\bo{k-q}}^{(n)*}(\bo{x})=f_{\bo{kq}}^{(n)}(\bo{x})\, .\lb{eq;Rsymmetry}
\ee
Then it is shown from Eq.~(\ref{def of Wannier}) that the Wannier
 functions become real. But, in our discussions below, we do not assume
the reflection symmetry. 

The field operator is expanded in terms of the set $\{w_{\bo{ki}}^{(n)}(\bo{x})\}$ as
\be
\hat{\phi}(\bo{x},t)=\sum_{n,\bo{i}}\hat{a}_{\bo{ki}}^{(n)}(t)w_{\bo{ki}}^{(n)}(\bo{x})\, , \lb{eq;FWanifield}
\ee
where the operator $\hat{a}_{\bo{ki}}^{(n)}$ satisfies the usual bosonic 
commutation relations,
$[\hat{a}_{\bo{ki}}^{(n)},\hat{a}_{\bo{kj}}^{(n')\dagger}]=\delta_{\bo{ij}}\delta_{nn'}$ and $[\hat{a}_{\bo{ki}}^{(n)},\hat{a}_{\bo{kj}}^{(n')}]=[\hat{a}_{\bo{ki}}^{(n)\dagger},\hat{a}_{\bo{kj}}^{(n')\dagger}]=0$. The solution
of the GP equation $v_{\bo{k}}(\bo{x})$ can be rewritten as
\be
v_{\bo{k}}(\bo{x})&=&n_{\rm c}^{\frac{1}{2}}\sum_{\bo{i}}e^{i\bo{k\cdot
x_i}}w_{\bo{ki}}^{(1)}(\bo{x}) \, , \lb{eq;FwaniGP}\\ 
n_{\rm c}&=&\frac{N_{\rm c}}{I_{\rm s}} \, .
\ee
Here we use Eq.~(\ref{zero mode}) and the inverse transformation of Eq.~(\ref{def of Wannier}),
\be
f_{\bo{kq}}^{(n)}(\bo{x})=I_{\rm s}^{-\frac{1}{2}}
\sum_{\bo{i}}e^{i\bo{q\cdot x_{\bo{i}}}} w_{\bo{ki}}^{(n)}(\bo{x}) \, ,
\label{eq:f-Wannier}
\ee
following from the formula,
\be
\frac{1}{I_{\rm s}} \sum_{\bo{i}}e^{i\bo{(q'-q)\cdot x_i}}=
\delta_{\bo{q'q}}\, .  
\ee
Substituting Eq.~(\ref{eq;FwaniGP}) into Eq.~(\ref{eq;GP}) and Eqs.~(\ref{eq;FWanifield}) and
(\ref{eq;FwaniGP}) into Eq.~(\ref{eq;hami}), and making use of Eq.~(\ref{eq:WaniOrthoNorm}),
one can rewrite  the GP equation and  the unperturbed Hamiltonian as 
\begin{widetext}
\be
0&=&\sum_{\bo{i}_{2}} \left\{ -J_{\bo{i}_{1}\bo{i}_{2}}^{(n1)}+n_{\rm c}
 \sum_{\bo{i}_3,\bo{i}_4} e^{i\bo{k}\cdot(\bo{x}_{\bo{i}_{4}} -
 \bo{x}_{\bo{i}_{3}})}
 U_{\bo{i}_{1}\bo{i}_{2}\bo{i}_{3}\bo{i}_{4}}^{(n111)}-\left(\mu 
 +\varepsilon\bar{\epsilon}\right)\delta_{\bo{i}_1 \bo{i}_2} \delta_{n1} 
 \right\}e^{i\bo{k}\cdot \bo{x}_{\bo{i}_{2}}}  \lb{eq;GP2} \, , \\ 
\hat{H}_{0}&=&\sum_{n_{1},n_{2},\bo{i}_1,\bo{i}_2}
\left\{ \left(-J_{\bo{i}_{1}
 \bo{i}_{2}}^{(n_{1}n_{2})}-\mu\delta_{\bo{i}_{1}\bo{i}_{2}}\delta_{n_{1}
 n_{2}}\right) 
\hat{a}_{\bo{k}\bo{i}_{1}}^{ (n_{1})
 \dag}\hat{a}_{\bo{k}\bo{i}_{2}}^{(n_{2}) } \right.\nonumber \\ 
& &{} +\frac{n_{\rm c}}{2} \sum_{\bo{i}_3,\bo{i}_4}\left(4 U_{\bo{i}_{1}
 \bo{i}_{2} \bo{i}_{3} \bo{i}_{4}}^{(n_{1}n_{2}11)}
e^{i\bo{k}\cdot(\bo{x}_{\bo{i}_{4}}-\bo{x}_{\bo{i}_{3}})}\hat{a}_{\bo{k}
 \bo{i}_{1}}^{(n_{1})\dag} \hat{a}_{\bo{k}\bo{i}_{2}}^{(n_{2})} 
+ U_{\bo{i}_{3} \bo{i}_{1} \bo{i}_{4}
 \bo{i}_{2}}^{(1n_{1}1n_{2})}e^{-i\bo{k} \cdot (\bo{x}_{\bo{i}_{4}} +
 \bo{x}_{\bo{i}_{3}})} \hat{a}_{\bo{k} \bo{i}_{1}}^{(n_{1})}
 \hat{a}_{\bo{k} \bo{i}_{2}}^{(n_{2})} \right. \nonumber \\ 
& & \left.\left.\qquad\qquad\qquad {} + U_{\bo{i}_{1} \bo{i}_{3}
 \bo{i}_{2} \bo{i}_{4}}^{(n_{1}1n_{2}1)}
e^{i\bo{k} \cdot ( \bo{x}_{\bo{i}_{4}} + \bo{x}_{\bo{i}_{3}})}
 \hat{a}_{\bo{k} \bo{i}_{1}}^{(n_{1})\dag} 
\hat{a}_{\bo{k} \bo{i}_{2}}^{(n_{2})\dag} \right) \right\} \,
 . \lb{eq:hami2} 
\ee
\end{widetext}
with the notations of  
\be
J_{\bo{i}_{1} \bo{i}_{2}}^{(n_{1}n_{2})}
&=&-\int \! d^{3} x \, w_{\bo{ki_{1}}}^{(n_{1})*}
\left(K+V_{\mathrm{opt}}\right) w_{\bo{ki_{2}}}^{(n_{2})} \, , \nonumber
\\ 
U_{\bo{i}_{1} \bo{i}_{2} \bo{i}_{3} \bo{i}_{4}}^{(n_{1}n_{2}n_{3}n_{4})}
&=& g \int \! d ^{3}x \, w_{\bo{k} \bo{i}_{1}}^{(n_{1})*} w_{\bo{k}
\bo{i}_{2}}^{(n_{2})} w_{\bo{k}\bo{i}_{3}}^{(n_3)*}
w_{\bo{k}\bo{i}_{4}}^{(n_4)}\, .\nonumber \\ 
\ee

Suppose that the system is so cold that we may restrict excitations only
in the first band. Then the completeness condition holds in the first band,
\be
& &\sum_{\bo{i}}w_{\bo{ki}}^{*}(\bo{x}')
w_{\bo{ki}}(\bo{x})=\delta(\bo{x}-\bo{x}') \,,  
\ee
or equivalently from  Eq.~(\ref{eq:f-Wannier}), 
\be 
& &\sum_{\bo{q}} f_{\bo{kq}}^{*}(\bo{x'}) f_{\bo{kq}}(\bo{x}) =
\delta(\bo{x}-\bo{x'})  \, . 
\ee 
Here and hereafter the band index ($n=1$) is omitted.

By taking the most relevant terms for the kinetic terms
$J_{\bo{i}_{1} \bo{i}_{2}}$ and the on-site terms for interaction
$U_{\bo{i}_{1} \bo{i}_{2} \bo{i}_{3} \bo{i}_{4}}$ (tight-binding limit), 
Eqs.~(\ref{eq;GP2}) and (\ref{eq:hami2}) are reduced to  
\be
z - \mu &=& \sum_{l=x,y,z} \left( J_{l}e^{ik_{i}d} + J_{l}^{*}
e^{-ik_{i}d}\right)- Un_{\rm c} + \varepsilon \bar{\epsilon} \, ,
\lb{eq;FCP} \\
\hat{H}_{0} &=& -\sum_{<\bo{i},\bo{j}>} J_{\bo{i}\bo{j}}
\hat{a}_{\bo{ki}}^{\dag} \hat{a}_{\bo{kj}}\nonumber \\ 
& & {} + \sum_{\bo{i}} \Big{\{} (z-\mu+2Un_{\rm c})
\hat{a}_{\bo{ki}}^{\dagger}\hat{a}_{\bo{ki}} \nonumber \\ 
& &{ } + \frac{Un_{\rm c}}{2} \left( e^{-2i\bo{k\cdot x_{\bo{i}}}}
\hat{a}_{\bo{ki}} \hat{a}_{\bo{ki}} + e^{2i\bo{k\cdot}\bo{x}_{\bo{i}}}
\hat{a}_{\bo{ki}}^{\dag} \hat{a}_{\bo{ki}}^{\dag} \right) \Big{\}}, \,
\lb{eq;FtightHami} \nonumber \\ 
\ee
where $<\bo{i},\bo{j}>$ represents the sum over the nearest neighbours 
and the following notations are introduced: 
$J_{l}=J_{\bo{ii}+\bo{1}_{l}}$, $J_{l}^{*}=J_{\bo{i}+\bo{1}_{l}\bo{i}}$,
$z=-J_{\bo{ii}}$ and $U=U_{\bo{iiii}}$.
Here $\bo{1}_{l}$ is the unit vector along the $l$-direction, e.g. 
$\bo{1}_{x}=(1,0,0)$. 
 We consider  the Fourier representation of the operators
 $\hat{a}_{\bo{ki}}$ and $\hat{a}_{\bo{ki}}^{\dag}$ 
\be
\hat{a}_{\bo{ki}} &=& I_{\rm s}^{-\frac{1}{2}}
\sum_{\bo{q}}\hat{c}_{\bo{kq}}e^{i\bo{q} \cdot \bo{x}_{\bo{i}}} \, ,
\label{eq:a-c}\\  
\hat{a}_{\bo{ki}}^{\dag} &=& I_{\rm s}^{-\frac{1}{2}} \sum_{\bo{q}}
\hat{c}_{\bo{kq}}^{\dag} e^{-i\bo{q} \cdot \bo{x_{\bo{i}}}} \, , 
\ee
where the operators $\hat{c}_{\bo{kq}}$ and
$\hat{c}_{\bo{kq}}^{\dag}$ satisfy the bosonic commutation relations 
\be
\big{[}\hat{c}_{\bo{kq}},\hat{c}_{\bo{kq'}}^{\dag} \big{]} =
\delta_{\bo{qq'}}\, , \\ 
\big{[}\hat{c}_{\bo{kq}},\hat{c}_{\bo{kq'}}\big{]} =
\big{[}\hat{c}_{\bo{kq}}^{\dag}, \hat{c}_{\bo{kq'}}^{\dag} \big{]} = 0
\, . 
\ee

Substituting the Fourier representation of the operators
$\hat{a}_{\bo{ki}}$, $\hat{a}_{\bo{ki}}^{\dag}$ and the chemical
potential (\ref{eq;FCP}) into the unperturbed Hamiltonian
(\ref{eq;FtightHami}), we reduce it to
\be
\hat{H}_{0} &=& \sum_{\bo{\bar{q}}} \lambda_{\bo{k\bar{q}}}
\hat{c}_{\bo{k\bar{q}}}^{\dag} \hat{c}_{\bo{k\bar{q}}} \nonumber \\  
& & {} +\frac{Un_{\rm c}}{2} \sum_{\bo{\bar{q}}} \left( 2
\hat{c}_{\bo{k\bar{q}}}^{\dag} \hat{c}_{\bo{k\bar{q}}} +
\hat{c}_{\bo{k}-\bo{\bar{q}}} \hat{c}_{\bo{k\bar{q}}} +
\hat{c}_{\bo{k}-\bo{\bar{q}}}^{\dag} \hat{c}_{\bo{k\bar{q}}}^{\dag}
\right) \, , \lb{eq;Fmainhami2}  \nonumber \\ 
\\
\lambda_{\bo{k\bar{q}}} &=& \sum_{l=x,y,z} 4 |J_{l}| \sin \left(
\Theta_{l} + k_{l} d + \frac{\bar{q}_{l}d}{2} \right)
\sin \left( \frac{\bar{q}_{l}d}{2} \right) + \varepsilon \bar{\epsilon}
\, , \nonumber \\ \label{eq:lambda}
\ee
where we write $J_{l}=|J_{l}|e^{i\Theta_{l}}$ and replace 
$\bo{q}$ with $\bo{k}+\bo{\bar{q}}$.
Hereafter our notation is simplified as
$\lambda_{\bo{kq}}=\lambda_{\bo{kk}+\bo{{\bar q}}} \rightarrow 
\lambda_{\bo{k{\bar q}}}$. 
Each component of $\bar{q}_{\bo{i}}$ runs only over
 $ -\pi/d \le \bar{q}_i \le \pi/d $ 
$(i=x,y,z)$ in the summation,
because $\hat{c}_{\bo{kq}}(t)$ and $\lambda_{\bo{kq}}$ have the
periodicity as  $\hat{c}_{\bo{kq}+\bo{K}}(t)=\hat{c}_{\bo{kq}}(t)$ and
$\lambda_{\bo{kq}+\bo{K}}=\lambda_{\bo{kq}}$  where $\bo{K}$ is the
reciprocal vector. 

The expansion of the quantum field operator Eq.~(\ref{eq;FWanifield})
 can be rewritten with help of Eqs.~(\ref{eq:f-Wannier}) and
 (\ref{eq:a-c}) as 
\be
\hat{\phi}(x)=\sum_{\bo{\bar{q}}}\hat{c}_{\bo{k\bar{q}}}(t) 
f_{\bo{k\bar{q}}}(\bo{x})
\lb{eq:phi-cf}
\ee
under the tight-binding approximation.


\section{Hamiltonian with complex eigenvalues and its eigenstates}
\label{sec:Hdiag}
Let us consider the  equation of time evolution for
$\hat{c}_{\bo{k\bar{q}}} (t)$  and $\hat{c}_{\bo{k}-\bo{\bar{q}}}^{\dag}
(t)$ as 
\be
i\hbar\frac{d}{dt}\hat{\bo{c}}_{\bo{k\bar{q}}}(t)= 
\left[\hat{\bo{c}}_{\bo{k\bar{q}}}(t),\hat{H}_{0}\right]
=T_{\bo{k\bar{q}}}\hat{\bo{c}}_{\bo{k\bar{q}}}(t)\, ,\lb{eq;Ftimeevo} 
\ee
where we have introduced the doublet notation 
\be
\hat{\bo{c}}_{\bo{k\bar{q}}}(t) =\left(
\begin{array}{c}
\hat{c}_{\bo{k\bar{q}}}(t) \\
\hat{c}_{\bo{k}-\bo{\bar{q}}}^{\dagger}(t)
\end{array}
\right). 
\ee
The $2\times 2$-matrix $T_{\bo{k\bar{q}}}$ is given by
\be
T_{\bo{k\bar{q}}}=\left(
\begin{array}{cc}
(\lambda_{\bo{k\bar{q}}}+Un_{\rm c}) & Un_{\rm c} \\
-Un_{\rm c} & -(\lambda_{\bo{k}-\bo{\bar{q}}}+Un_{\rm c})
\end{array}
\right) \, .
\ee
To solve Eq.~(\ref{eq;Ftimeevo}), we attempt
to diagonalize $T_{\bo{k\bar{q}}}$. Let us set up the eigenequation 
\be
T_{\bo{k\bar{q}}}\bo{x}_{\bo{k\bar{q}}} = \hbar \omega_{\bo{k\bar{q}}}
\bo{x}_{\bo{k\bar{q}}} \, , \lb{eq:FC1} 
\ee
which gives two independent eigenvalues
$\hbar\omega_{\bo{k\bar{q}}}^{(\pm)}$ and doublet eigenvectors
$\bo{x}_{\bo{k\bar{q}}}^{(\pm)}$. 
The eigenvalues $\hbar\omega_{\bo{k\bar{q}}}^{(\pm)}$ are found to be 
\be
\hbar \omega_{\bo{k\bar{q}}}^{(\pm)} &=& \hbar
\omega_{\bo{k\bar{q}}}^{(1)} \pm \hbar \omega_{\bo{k\bar{q}}}^{(2)} \, , 
\lb{eq:hbaromegapm} \\
\hbar \omega_{\bo{k\bar{q}}}^{(1)} &=& \frac{1}{2} (
\lambda_{\bo{k\bar{q}}} - \lambda_{\bo{k}-\bo{\bar{q}}}) \, , \\
\hbar \omega_{\bo{k\bar{q}}}^{(2)} &=& \sqrt{\Lambda_{\bo{k\bar{q}}}^{2}
+ 2 Un_{\rm c} \Lambda_{\bo{k\bar{q}}}}\, , \lb{eq:hbaromega2}
\ee
where 
\be
\Lambda_{\bo{k\bar{q}}} = \frac{\lambda_{\bo{k\bar{q}}} +
\lambda_{\bo{k}-\bo{\bar{q}}}}{2}\, . \lb{eq:Lambda}
\ee

\begin{figure}[htb]
\begin{center}
\includegraphics[width=1.0\linewidth]{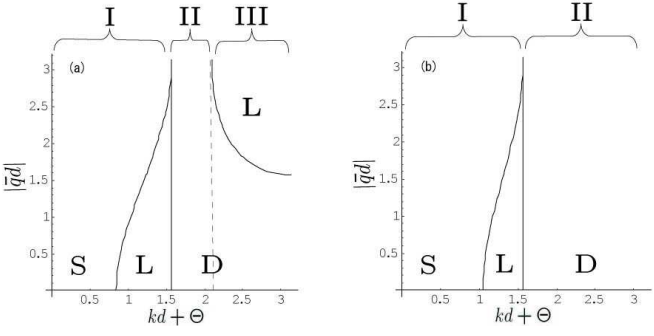}
\end{center}
\caption{\footnotesize{
Stability phase diagrams for (a) $n_{\rm c} U /|J|=1$, and (b) $n_{\rm
 c} U /|J|=2$ in 1D system. 
The regions {\bf S}, {\bf L} and {\bf D} represent those in which the
 eigenvalue $\hbar\omega_{\bo{k\bar{q}}}^{(+)}$ is positive, negative
 and complex, respectively. We classify the axis of abscissas into three
 regions, denoted by \Roman{ichi}, \Roman{ni} and \Roman{san}. The
 eigenvalue is always complex in the region \Roman{ni}, while it is
 always real in the region \Roman{ichi}. In the region \Roman{san}, it
 can be real or complex.  
}}
\label{Stability Phase Diagram}
\end{figure}

Note that these eigenvalues become complex if the condition, 
\be
\Lambda_{\bo{k\bar{q}}}^{2} + 2Un_{\rm c} \Lambda_{\bo{k\bar{q}}} < 0 \,
, 
\ee
is satisfied. 
In Fig. \ref{Stability Phase Diagram}, we plot the stability phase
diagram for different values of $n_{\rm c} U /|J|$. 

For a one-dimensional system, this condition is rewritten simply as
\be
\cos(kd+\Theta) \left\{ \cos(kd+\Theta) \sin^{2} \left(
\frac{\bar{q}d}{2} \right) + \frac{Un_{\rm c}}{2J} \right\} < 0
\ee
at the limit of $\varepsilon \to 0$.

\subsection{Some properties of eigenvalues and eigenvectors}

We investigate some common properties of the eigenvalues
$\hbar\omega_{\bo{k\bar{q}}}^{(\pm)}$ and eigenvectors
$\bo{x}_{\bo{k\bar{q}}}^{(\pm)}$. 

 We find the following two algebraic properties of the non-Hermitian
 matrix $T_{\bo{k\bar{q}}}$ which may be expressed as
\be
T_{\bo{k\bar{q}}} = \hbar \omega_{\bo{k\bar{q}}}^{(1)} \sigma_0 +
 (\Lambda_{\bo{k\bar{q}}} + Un_{\rm c}) \sigma_3 + i Un_{\rm c} \sigma_2 
\, ,
\ee
where $\sigma_0$ is the unit matrix and $\sigma_{i}$ $(i=1,2,3)$
 represents $i$-th Pauli matrix. 

As the first property, the matrix $T_{\bo{k\bar{q}}}$ is
pseudo-Hermitian: 
\be
\sigma_3 T^\dag_{\bo{k\bar{q}}} \sigma_3 = T_{\bo{k\bar{q}}} \, .
\label{eq:pseudT}
\ee
Then the inner-product of the doublet eigenvectors is naturally
introduced as 
\be
(\bo{x},\bo{x}') = \bo{x}^\dag \sigma_3 \bo{x}' \, ,
\ee
since we will later see the orthogonality of eigenvectors with respect to 
this inner-product, coming from the relation
\be
(\bo{x},T\bo{x}')=(T\bo{x},\bo{x}') \, , \label{eq:xTx}
\ee
for any pseudo-Hermitian matrix $T$ in the sense of Eq.~(\ref{eq:pseudT}).
Note that the metric with respect to this inner-product is indefinite.

The second property is derived as 
\be
\sigma_{1} T_{\bo{k\bar{q}}} \sigma_{1} &=& 
\hbar \omega_{\bo{k\bar{q}}}^{(1)} \sigma_0 -(\Lambda_{\bo{k\bar{q}}} +
Un_{\rm c}) \sigma_3  - i Un_{\rm c} \sigma_2 \nonumber \\
&=& - \hbar \omega_{\bo{k}-\bo{\bar{q}}}^{(1)} \sigma_0 - (
\Lambda_{\bo{k} - \bo{\bar{q}}} + Un_{\rm c} ) \sigma_3  
-i Un_{\rm c} \sigma_2 \nonumber \\
&=& - T_{\bo{k}-\bo{\bar{q}}} \, .
\label{eq:sigma1T}
\ee
It is also clear from Eqs.~(\ref{eq:hbaromegapm})--(\ref{eq:Lambda})
that the eigenvalues
$\hbar\omega_{\bo{k\bar{q}}}^{(\pm)}$ and
$\hbar\omega_{\bo{k}-\bo{\bar{q}}}^{(\pm)}$ are related to each other,
\be
\hbar \omega_{\bo{k\bar{q}}}^{(\pm)} = - \hbar
\omega_{\bo{k}-\bo{\bar{q}}}^{(\mp)}\, . \lb{eq:+-} 
\ee
Thus the eigenequations
 $T_{\bo{k}-\bo{\bar{q}}}\bo{x}_{\bo{k}-\bo{\bar{q}}}^{(\pm)} 
= \hbar \omega_{\bo{k}-\bo{\bar{q}}}^{(\pm)}
\bo{x}_{\bo{k}-\bo{\bar{q}}}^{(\pm)}$ imply $T_{\bo{k\bar{q}}}
\sigma_1 \bo{x}_{\bo{k}-\bo{\bar{q}}}^{(\pm)} = \hbar
\omega_{\bo{k\bar{q}}}^{(\mp)} \sigma_1
\bo{x}_{\bo{k}-\bo{\bar{q}}}^{(\pm)}$, and therefore
\be
\bo{x}_{\bo{k\bar{q}}}^{(\mp)} = \sigma_{1} \bo{x}_{\bo{k} -
\bo{\bar{q}}}^{(\pm)} \, .\lb{eq:Asyn2} 
\ee

We may introduce the traceless part of $T_{\bo{k\bar{q}}}$ by
\be
{\tilde T}_{\bo{k\bar{q}}} = ( \Lambda_{\bo{k\bar{q}}}+Un_{\rm c})
\sigma_3 + i Un_{\rm c} \sigma_2 \, ,
\ee
for which we have the eigenequations
\be
{\tilde T}_{\bo{k\bar{q}}} \bo{x}_{\bo{k\bar{q}}}^{(\pm)} = 
\pm \hbar \omega_{\bo{k\bar{q}}}^{(2)}\bo{x}_{\bo{k\bar{q}}}^{(\pm)} \,
, 
\ee
where $\hbar \omega_{\bo{k\bar{q}}}^{(2)}$ is defined in
Eq.~(\ref{eq:hbaromega2}).  Note that $\bo{x}_{\bo{k\bar{q}}}^{(\pm)}$
and $\bo{x}_{\bo{k-\bar{q}}}^{(\pm)}$ are degenerate states because 
$\hbar \omega_{\bo{k\bar{q}}}^{(2)} = \hbar
\omega_{\bo{k-\bar{q}}}^{(2)}$. 
The matrix ${\tilde T}_{\bo{k\bar{q}}}$ satisfies the algebraic relation, 
\be
\sigma_{1} {\tilde T}_{\bo{k\bar{q}}} \sigma_{1} = - {\tilde
T}_{\bo{k\bar{q}}} \, .  \label{eq:sigma1tilT}
\ee
Similarly in the previous paragraph, we easily find that
\be
{\tilde T}_{\bo{k\bar{q}}} \sigma_1 \bo{x}_{\bo{k\bar{q}}}^{(\pm)} = 
\mp \hbar \omega_{\bo{k\bar{q}}}^{(2)}
\sigma_1\bo{x}_{\bo{k\bar{q}}}^{(\pm)} 
\ee
leading to
\be
\bo{x}_{\bo{k\bar{q}}}^{(\mp)} = \sigma_{1}
\bo{x}_{\bo{k\bar{q}}}^{(\pm)} \, . \lb{eq:Asyn3} 
\ee

Thus we can express the four eigenvectors
$\bo{x}_{\bo{k\bar{q}}}^{(\pm)}$ and
$\bo{x}_{\bo{k}-\bo{\bar{q}}}^{(\pm)}$ only by two parameters, say
$u_{\bo{k\bar{q}}}$ and $v_{\bo{k\bar{q}}}$, except for a trivial phase
factor: 
\be
\bo{x}_{\bo{k\bar{q}}}^{(+)}= \bo{x}_{\bo{k}-\bo{\bar{q}}}^{(+)}&=&
\left(
\begin{array}{c}
u_{\bo{k\bar{q}}} \\
v_{\bo{k\bar{q}}}
\end{array}
\right) ,\lb{eq:xuv+} \\
\bo{x}_{\bo{k\bar{q}}}^{(-)}= \bo{x}_{\bo{k}-\bo{\bar{q}}}^{(-)} &=&
\left( 
\begin{array}{c}
v_{\bo{k\bar{q}}}  \\
u_{\bo{k\bar{q}}}
\end{array}
\right) .
\lb{eq:xuv-}
\ee

\subsection{Real eigenvalues}

In this subsection we consider the case where all eigenvalues are real,
e.g., the region \Roman{ichi} in Fig.~1.
From Eq.~(\ref{eq:xTx}), one obtains the following relation
\be
(\hbar\omega_{\bo{k\bar{q}}}^{(s')}-\hbar\omega_{\bo{k\bar{q}}}^{(s)})
(\bo{x}_{\bo{k\bar{q}}}^{(s)},\bo{x}_{\bo{k\bar{q}}}^{(s')})=0 \, ,
\lb{eq:orthogonality}
\ee
where $s$ and $s'$ are $+$ or $-$.
Therefore we obtain the orthogonality condition 
\be
(\bo{x}_{\bo{k\bar{q}}}^{(-)},\bo{x}_{\bo{k\bar{q}}}^{(+)})
=(\bo{x}_{\bo{k\bar{q}}}^{(+)},\bo{x}_{\bo{k\bar{q}}}^{(-)})
=0 \, . \lb{eq:oth3}
\ee 
Noting that $u_{\bo{k\bar{q}}}$ and $v_{\bo{k\bar{q}}}$ are real numbers,
we may set the normalization condition as
\be
u_{\bo{k\bar{q}}}^{2}-v_{\bo{k\bar{q}}}^{2}&=&1\, . \lb{eq:bogo}
\ee
Then Eqs.~(\ref{eq:xuv+}) and (\ref{eq:xuv-}) 
imply that the eigenvectors are normalized with the indefinite metric as
\be
(\bo{x}_{\bo{k\bar{q}}}^{(+)},\bo{x}_{\bo{k\bar{q}}}^{(+)})&=&1 \,
,\lb{eq:oth1}\\ 
(\bo{x}_{\bo{k\bar{q}}}^{(-)},\bo{x}_{\bo{k\bar{q}}}^{(-)})&=&-1 \,
.\lb{eq:oth2} 
\ee

The elements of the eigenvectors are determined explicitly as 
\be
u_{\bo{k\bar{q}}} &=& \sqrt{\frac{1}{2} \left( 1 +
\frac{\Lambda_{\bo{k\bar{q}}} + Un_{\rm c}}{\hbar
\omega_{\bo{k\bar{q}}}^{(2)}}\right)} \, , \lb{eq:vec1} \\ 
v_{\bo{k\bar{q}}} &=& - \sqrt{\frac{1}{2} \left( -1 +
\frac{\Lambda_{\bo{k\bar{q}}} + Un_{\rm c}}{\hbar
\omega_{\bo{k\bar{q}}}^{(2)}}\right)}  \, , \lb{eq:vec2}  
\ee 
Note that $u_{\bo{k\bar{0}}}$ and $v_{\bo{k\bar{0}}}$ are singular
in the limit of $\varepsilon \to 0$.

The matrix $T_{\bo{k\bar{q}}}$ is diagonalized by the matrix
$O_{\bo{k\bar{q}}}$ as 
\be
O_{\bo{k\bar{q}}}^{-1}T_{\bo{k\bar{q}}}O_{\bo{k\bar{q}}}=\left(
\begin{array}{cc}
\hbar\omega_{\bo{k\bar{q}}}^{(+)} &0 \\
0& \hbar\omega_{\bo{k\bar{q}}}^{(-)}
\end{array}
\right)\, ,
\ee
where
\be 
O_{\bo{k\bar{q}}}&=&\left(
\begin{array}{cc}
\bo{x}_{\bo{k\bar{q}}}^{(+)} &\bo{x}_{\bo{k\bar{q}}}^{(-)} 
\end{array}
\right) \, .
\ee
The orthonormal conditions (\ref{eq:oth3}), (\ref{eq:oth1})
 and (\ref{eq:oth2}) 
are equivalent 
to $O_{\bo{k\bar{q}}}^\dag \sigma_3 O_{\bo{k\bar{q}}} = \sigma_3$,
yielding 
\be
O_{\bo{k\bar{q}}}^{-1}= 
\sigma_{3}O^\dagger_{\bo{k\bar{q}}}\sigma_3\, . \lb{eq:real relation}
\ee
The operators, defined by
\be
\hat{\bo{b}}_{\bo{k\bar{q}}} =\left(
\begin{array}{c}
\hat{b}_{\bo{k\bar{q}}}  \\
\hat{b}_{\bo{k}-\bo{\bar{q}}}^{\dagger}
\end{array}
\right)&=&O_{\bo{k\bar{q}}}^{-1}\hat{\bo{c}}_{\bo{k\bar{q}}}
\nonumber \\
&=&\left(
\begin{array}{cc}
u_{\bo{k\bar{q}}} & -v_{\bo{k\bar{q}}} \\
-v_{\bo{k\bar{q}}}& u_{\bo{k\bar{q}}}
\end{array}
\right) \left(
\begin{array}{c}
\hat{c}_{\bo{k\bar{q}}}  \\
\hat{c}_{\bo{k}-\bo{\bar{q}}}^{\dagger}
\end{array}
\right) \lb{eq:transR} \, , \nonumber \\
\ee
satisfy the  bosonic commutation relations
$[\hat{b}_{\bo{k\bar{q}}},\hat{b}_{\bo{k\bar{q}'}}^{\dagger}] =
\delta_{\bo{\bar{q}\bar{q}'}}$ and $[\hat{b}_{\bo{k\bar{q}}} ,
\hat{b}_{\bo{k\bar{q}'}}]=0$.
By using the $\hat b$-operators, one obtains the equation of 
time evolution in a diagonalized form from Eq.~(\ref{eq;Ftimeevo}),
\be
i\hbar\frac{d}{dt}\hat{\bo{b}}_{\bo{k\bar{q}}}(t)= \left(
\begin{array}{cc}
\hbar \omega_{\bo{k\bar{q}}}^{(+)} & 0 \\
 0& -\hbar \omega_{\bo{k}-\bo{\bar{q}}}^{(+)}
\end{array}
\right)\hat{\bo{b}}_{\bo{k\bar{q}}}(t)
\, ,\lb{eq;Ftimeevob} 
\ee
where Eq.~(\ref{eq:+-}) has been made use of. 
Thus the time evolution of the operator $\hat{b}_{\bo{k\bar{q}}}$ is
simply given as 
\be
\hat{b}_{\bo{k\bar{q}}}(t) = \hat{b}_{\bo{k\bar{q}}}
e^{-i\omega_{\bo{k\bar{q}}}^{(+)}t} \, .  
\ee
In terms of the ${\hat b}$-operators,
the unperturbed Hamiltonian (\ref{eq;Fmainhami2}) is diagonalized as
\be
\hat{H}_{0}=\sum_{\bo{\bar{q}}} \hbar \omega_{\bo{k\bar{q}}}^{(+)}
\hat{b}_{\bo{k\bar{q}}}^{\dag} \hat{b}_{\bo{k\bar{q}}} + \rm{const}. 
\ee
The real eigenvalue $\hbar\omega_{\bo{k\bar{q}}}^{(+)}$ can be
interpreted as an energy of the quasi-particle, and can be negative in
our present model. The negative energy causes an energetic instability
which is called the Landau instability. 
For one-dimensional system, the condition of the  quasi-particle energy
$\hbar\omega_{k\bar{q}}^{(+)}$ being negative is written as
\be
& & \cos^2 \left( \frac{\bar{q}d}{2} \right) > \cos(kd+\Theta) \left\{
\cos(kd+\Theta) + \frac{Un_{\rm c}}{2J} \right\} \, , \nonumber \\ 
& &  \sin(kd+\Theta)\sin(\bar{q}d) <0
\ee
at the limit of $\varepsilon \to 0$.

The operator $\hat{b}_{\bo{k\bar{0}}}$ represents the zero-mode, as
its energy eigenvalue becomes
\be
\hbar \omega_{\bo{k\bar{0}}} = \left(\varepsilon \bar{\epsilon}
\right)^{\frac{1}{2}} \left( 2 Un_{\rm c} + \varepsilon \bar{\epsilon}
\right)^{\frac{1}{2}} \rightarrow 0 \quad \mbox{as} \quad \varepsilon
\rightarrow 0 \, , 
\ee
from Eqs.~(\ref{eq:lambda}), (\ref{eq:hbaromegapm})--(\ref{eq:Lambda}).
This zero-mode is presumed to be the Nambu-Goldstone (NG) mode 
appearing in the spontaneous breakdown of
global phase symmetry.  In order to prove that it is actually
the NG mode, one needs to check the Ward-Takahashi relations \cite{Enomoto},
but this has not been confirmed yet when complex modes appear.
The singular elements of $\bo{x}_{\bo{k\bar{0}}}^{(\pm)}$ are
calculated from
Eqs.~(\ref{eq:vec1}) and (\ref{eq:vec2}) as
\be
u_{\bo{k\bar{0}}} &=& ( 4 \varepsilon\bar{\epsilon})^{-\frac{1}{4}}
\left\{ \alpha + \frac{ (
\varepsilon\bar{\epsilon})^{\frac{1}{2}}}{2\alpha}\right\} + O \left(
\varepsilon^{\frac{3}{4}} \right) \, , \\ 
v_{\bo{k\bar{0}}} &=& -( 4 \varepsilon\bar{\epsilon})^{-\frac{1}{4}}
\left\{ \alpha-\frac{(\varepsilon\bar{\epsilon})^{\frac{1}{2}}}{2\alpha}
\right\}+O \left( \varepsilon^{\frac{3}{4}} \right) \, , 
\ee
where $\alpha=\left(Un_{\rm c}/2 \right)^{1/4}$.
Regularizing the zero-mode this way, we can include it as a real mode,
as was done in our previous works \cite{Okumura1,MOY}. 

\subsection{Complex eigenvalues}

Next, let us consider the case where all eigenvalues are complex, e.g.,
the region \Roman{ni} in Fig.~1.  
Then the eigenvalues $\hbar\omega_{\bo{k\bar{q}}}^{(\pm)}$ are complex
conjugate to each other as
$\hbar\omega_{\bo{k\bar{q}}}^{(-)}=\hbar\omega_{\bo{k\bar{q}}}^{(+)*}$.  
Now the elements of eigenvector $u_{\bo{k\bar{q}}}$ and
$v_{\bo{k\bar{q}}}$ become complex numbers. 
From the relation $(\hbar\omega_{\bo{k\bar{q}}}^{(s')*} -
\hbar\omega_{\bo{k\bar{q}}}^{(s)}) (\bo{x}_{\bo{k\bar{q}}}^{(s')},
\bo{x}_{\bo{k\bar{q}}}^{(s)}) =0$ where the superscript $s$ stands for
$\pm$, we obtain 
\be
(\bo{x}_{\bo{k\bar{q}}}^{(+)},\bo{x}_{\bo{k\bar{q}}}^{(+)})
=(\bo{x}_{\bo{k\bar{q}}}^{(-)},\bo{x}_{\bo{k\bar{q}}}^{(-)})&=&0\, ,
\lb{eq:n1}\\
(\bo{x}_{\bo{k\bar{q}}}^{(-)},\bo{x}_{\bo{k\bar{q}}}^{(+)})&=&C \, ,
\lb{eq:Const} 
\ee 
where Eqs.~(\ref{eq:xuv+}) and (\ref{eq:xuv-}) have been used and $C$ 
turns out to be a pure imaginary constant.
In order to fix the constant $C$ in Eq.~(\ref{eq:Const}), we take 
$u_{\bo{k\bar{q}}}^{2}-v_{\bo{k\bar{q}}}^{2}=1$ for convenience. Then
the expressions for the elements of the eigenvectors in
Eqs.~(\ref{eq:vec1}) and (\ref{eq:vec2}) are true for pure imaginary
$\hbar\omega_{\bo{k\bar{q}}}^{(2)}$, and the simple relation,
\be
v_{\bo{k\bar{q}}}^{*}=iu_{\bo{k\bar{q}}} \, ,\lb{eq:uvrelation}
\ee
is found. This choice corresponds to fixing $C=i$, and 
the orthonormal conditions (\ref{eq:n1}) and (\ref{eq:Const}) are summarized as
\be
O_{\bo{k\bar{q}}}^{\dagger}\sigma_{3}O_{\bo{k\bar{q}}}=\sigma_{2}\, ,\lb{eq:complex relation1}
\ee
which derives
\be
O_{\bo{k\bar{q}}}^{-1}= \sigma_{2}O_{\bo{k\bar{q}}}^{\dagger}\sigma_{3} \, ,\lb{eq:complex relation2}
\ee

Similarly as in Eq~(\ref{eq:transR}), we introduce the new operators,
\be
\bo{A}_{\bo{k\bar{q}}}= \left(
\begin{array}{c}
\hat{A}_{\bo{k\bar{q}}}  \\
\hat{B}_{\bo{k}-\bo{\bar{q}}}^{\dagger}
\end{array}
\right)&=&O_{\bo{k\bar{q}}}^{-1} \bo{c}_{\bo{k\bar{q}}}
\nonumber \\
&=&\left(
\begin{array}{cc}
u_{\bo{k\bar{q}}} &iu_{\bo{k\bar{q}}}^{*} \\
iu_{\bo{k\bar{q}}}^{*}& u_{\bo{k\bar{q}}}
\end{array}
\right) \left(
\begin{array}{c}
\hat{c}_{\bo{k\bar{q}}}  \\
\hat{c}_{\bo{k}-\bo{\bar{q}}}^{\dagger}
\end{array}
\right) . \lb{eq:transC} \nonumber \\
\ee
The operators $\hat{A}_{\bo{k\bar{q}}}$ and $\hat{B}_{\bo{k\bar{q}}}$
satisfy the following relations
\be
\hat{A}_{\bo{k\bar{q}}}&=&i\hat{A}_{\bo{k}-\bo{\bar{q}}}^{\dagger}\, , \\
\hat{B}_{\bo{k\bar{q}}}&=&-i\hat{B}_{\bo{k}-\bo{\bar{q}}}^{\dagger}\, .
\ee 
The commutation relations among the $\hat{A}_{\bo{k\bar{q}}}$
 and $\hat{B}_{\bo{k\bar{q}}}$
operators become 
\be
\big{[}\hat{A}_{\bo{k\bar{q}}},\hat{B}_{\bo{k\bar{q}'}}^{\dagger}\big{]}
=-\big{[}\hat{A}_{\bo{k\bar{q}}}^{\dagger},\hat{B}_{\bo{k\bar{q}'}}\big{]}
&=&\delta_{\bo{\bar{q}\bar{q}'}} \, , \lb{eq;CCCR1}\\
 \big{[}\hat{A}_{\bo{k\bar{q}}},\hat{B}_{\bo{k\bar{q}'}}\big{]}=
\big{[}\hat{A}_{\bo{k\bar{q}}}^\dagger,\hat{B}_{\bo{k\bar{q}'}}^\dagger\big{]}
&=&-i\delta_{\bo{\bar{q}}-\bo{\bar{q}'}} \, , \lb{eq;CCCR2}\\
  \big{[}\hat{A}_{\bo{k\bar{q}}},\hat{A}_{\bo{k\bar{q}'}}^{\dagger}\big{]}
=\big{[}\hat{B}_{\bo{k\bar{q}}},\hat{B}_{\bo{k\bar{q}'}}^{\dagger}\big{]}&=&0\, .\lb{eq;CCCR3}
\ee
The  equation of time evolution (\ref{eq;Ftimeevo}) is reduced to the
 diagonalized form,
\be
i\hbar\frac{d}{dt}\hat{\bo{A}}_{\bo{k\bar{q}}}(t)=
 \left(
\begin{array}{cc}
\hbar \omega_{\bo{k\bar{q}}}^{(+)} & 0 \\
 0& \hbar \omega_{\bo{k\bar{q}}}^{(+)*}
\end{array}
\right)\hat{\bo{A}}_{\bo{k\bar{q}}}(t)
\, ,\lb{eq;FtimeevoA} 
\ee
so the operators $\hat{A}_{\bo{k\bar{q}}}$ and
$\hat{B}_{\bo{k}-\bo{\bar{q}}}^{\dagger}$ develop in time with the
complex frequencies as follows:
\be
\hat{A}_{\bo{k\bar{q}}}(t) & = & e^{-i\omega_{\bo{k\bar{q}}}^{(+)}t}
\hat{A}_{\bo{k\bar{q}}} \, , \\
\hat{B}_{\bo{k}-\bo{\bar{q}}}^{\dag}(t) & = &
e^{-i\omega_{\bo{k\bar{q}}}^{(+)*}t}
\hat{B}_{\bo{k}-\bo{\bar{q}}}^{\dag} \, . 
\ee
The unperturbed Hamiltonian (\ref{eq;Fmainhami2}) 
is reduced to 
\be
\hat{H}_{0}=\sum_{\bo{\bar{q}}} \left(
\frac{\hbar\omega_{\bo{k\bar{q}}}^{(+)}}{2}
\hat{B}_{\bo{k\bar{q}}}^{\dag} \hat{A}_{\bo{k\bar{q}}}
+ \frac{\hbar\omega_{\bo{k\bar{q}}}^{(+)*}}{2}
\hat{A}_{\bo{k\bar{q}}}^{\dag} \hat{B}_{\bo{k\bar{q}}} \right) \, . 
\ee
This way the Hamiltonian for complex modes is put into a diagonal form,
but does not have a representation in a Fock space,
and the complex eigenvalue can not be interpreted as a  quasi-particle
energy.

\subsection{Hamiltonian and canonical commutation relations}

Generally both real and complex eigenvalues can coexist, e.g., the
region \Roman{san} in Fig.~1. 
In such a case, the arguments in the preceding two subsections give the
following Hamiltonian in a diagonal form,
\be
\hat{H}_{0}&=&\sum_{\bo{\bar{q}_{\rm c}}} \left(
\frac{\hbar\omega_{\bo{k\bar{q}_{\rm c}}}^{(+)}}{2}
\hat{B}_{\bo{k\bar{q}_{\rm c}}}^{\dag} \hat{A}_{\bo{k\bar{q}_{\rm c}}}
+ \frac{\hbar\omega_{\bo{k\bar{q}_{\rm c}}}^{(+)*}}{2}
\hat{A}_{\bo{k\bar{q}_{\rm c}}}^{\dag} \hat{B}_{\bo{k\bar{q}_{\rm c}}}
\right) \nonumber \\ 
& & \hspace{1cm} {} + \sum_{\bo{\bar{q}_{\rm r}}} \hbar
\omega_{\bo{k\bar{q}_{\rm r}}}^{(+)} \hat{b}_{\bo{k\bar{q}_{\rm
r}}}^{\dag} \hat{b}_{\bo{k\bar{q}_{\rm r}}}+\rm{const}.,
\label{eq:generalH0}
\ee
where the Bloch wave number $\bo{\bar{q}}$ is distinguished by its
subscript depending on the property of the eigenvalue, i.e.,
$\bo{\bar{q}_{\rm r}}$ for real eigenvalue and $\bo{\bar{q}_{\rm c}}$
for complex one.

The operator $\hat{c}_{\bo{k\bar{q}}}$ is now written in terms of
$\hat{b}_{\bo{k\bar{q}}}$, $\hat{A}_{\bo{k\bar{q}}}$ and
$\hat{B}_{\bo{k}-\bo{\bar{q}}}^{\dagger}$ as 
\be
\hat{c}_{\bo{k\bar{q}_{\rm r}}}(t) &=& u_{\bo{k\bar{q}_{\rm r}}}
\hat{b}_{\bo{k\bar{q}_{\rm r}}} e^{-i\omega_{\bo{k\bar{q}_{\rm
r}}}^{(+)}t} + v_{\bo{k\bar{q}_{\rm r}}}
\hat{b}_{\bo{k}-\bo{\bar{q}_{\rm r}}}^{\dag} 
e^{i\omega_{\bo{k}-\bo{\bar{q}_{\rm r}}}^{(+)}t} \, , \nonumber\\
\hat{c}_{\bo{k\bar{q}_{\rm c}}}(t) &=& u_{\bo{k\bar{q}_{\rm c}}}
\hat{A}_{\bo{k\bar{q}_{\rm c}}} e^{-i\omega_{\bo{k\bar{q}_{\rm
c}}}^{(+)}t} - i u^{*}_{\bo{k\bar{q}_{\rm c}}}
\hat{B}_{\bo{k}-\bo{\bar{q}_{\rm c}}}^{\dag}
e^{-i\omega_{\bo{k\bar{q}_{\rm c}}}^{(+)*}t} \nonumber \, .
\ee
The expansion of the field operator Eq.~(\ref{eq:phi-cf}) becomes
\begin{widetext}
\be
\hat{\phi}(x) &=& \sum_{\bo{\bar{q}_{\rm r}}} \left(
 u_{\bo{k\bar{q}_{\rm r}}} f_{\bo{k\bar{q}_{\rm r}}}(\bo{x})
\hat{b}_{\bo{k\bar{q}_{\rm r}}} e^{-i\omega_{\bo{k\bar{q}_{\rm
 r}}}^{(+)}t} + v_{\bo{k\bar{q}_{\rm r}}} f_{\bo{k\bar{q}_{\rm r}}}
 (\bo{x}) \hat{b}_{\bo{k}-\bo{\bar{q}_{\rm r}}}^{\dag}
 e^{i\omega_{\bo{k}-\bo{\bar{q}_{\rm r}}}^{(+)}t} \right) \nonumber \\
& & {} + \sum_{\bo{\bar{q}_{\rm c}}}\left( u_{\bo{k\bar{q}_{\rm c}}}
 f_{\bo{k\bar{q}_{\rm c}}}(\bo{x})\hat{A}_{\bo{k\bar{q}_{\rm c}}}
 e^{-i\omega_{\bo{k\bar{q}_{\rm c}}}^{(+)}t} - i
 u^{*}_{\bo{k\bar{q}_{\rm c}}} f_{\bo{k\bar{q}_{\rm c}}}(\bo{x})
\hat{B}_{\bo{k}-\bo{\bar{q}_{\rm c}}}^{\dag}
 e^{-i\omega_{\bo{k\bar{q}_{\rm c}}}^{(+)*}t} \right) \,
 . \lb{eq:fieldexpre} 
\ee
\end{widetext}
One can easily check that the field operators satisfy the CCRs in the
first band, 
\be
\big{[}\hat{\phi}(\bo{x},t), \hat{\phi}^{\dag}(\bo{x}',t)\big{]} &=& 
\delta(\bo{x}-\bo{x}')\, ,  \nonumber\\ 
\big{[}\hat{\phi}(\bo{x},t),\hat{\phi}(\bo{x}',t)\big{]} &=&
\big{[}\hat{\phi}^{\dag}(\bo{x},t), \hat{\phi}^{\dag}(\bo{x}',t) \big{]}
= 0 \, .
\ee

\subsection{Eigenstate of complex mode}

In order to find eigenstates of the Hamiltonian (\ref{eq:generalH0})
we introduce the operator that transforms
$\hat{c}_{\bo{k\bar{q}}}$ to $\hat{b}_{\bo{k\bar{q}}}$
or $\hat{A}_{\bo{k\bar{q}}}$ and $\hat{B}_{\bo{k\bar{q}}}$, defined by
\be
\hat{V}=\exp\Big{[}\sum_{\bo{\bar{q}}}
i\theta_{\bo{k\bar{q}}}\hat{G}_{\bo{k\bar{q}}}\Big{]} \, ,
\ee
where $\hat{G}_{\bo{k\bar{q}}}$ is
\be
\hat{G}_{\bo{k\bar{q}}} = i \left(\hat{c}_{\bo{k\bar{q}}}
\hat{c}_{\bo{k}-\bo{\bar{q}}} - \hat{c}_{\bo{k\bar{q}}}^{\dag}
\hat{c}_{\bo{k}-\bo{\bar{q}}}^{\dag} \right) \, .
\ee
Note that $\theta_{\bo{k\bar{q}}}$ is in general a complex number, and
it is written by a pair of
 real numbers, $\theta^{\rm R}_{\bo{k\bar{q}}}$ and
$\theta^{\rm I}_{\bo{k\bar{q}}}$, as 
\be
\theta_{\bo{k\bar{q}}}=\theta_{\bo{k\bar{q}}}^{\rm
R}+i\theta_{\bo{k\bar{q}}}^{\rm I} \, ,
\ee
so the operator $\hat{V}$ is not unitary unless the parameter
$\theta_{\bo{k\bar{q}}}^{\rm I}$ vanishes identically.
As is well known, the operator $\hat{V}$ gives rise to the Bogoliubov
transformation when all the parameters $\theta_{\bo{k\bar{q}}}$ 
are real \cite{Umezawa}.  Even for complex $\theta_{\bo{k\bar{q}}}$, 
the transformation generated by the operator $\hat{V}$ may be written
down 
as
\be
\hat{V}\left(
\begin{array}{c}
\hat{c}_{\bo{k\bar{q}}}  \\
\hat{c}_{\bo{k}-\bo{\bar{q}}}^{\dagger}
\end{array}
\right)\hat{V}^{-1}
= O^{-1}(\theta_{\bo{k\bar{q}}})\left(
\begin{array}{c}
\hat{c}_{\bo{k\bar{q}}}  \\
\hat{c}_{\bo{k}-\bo{\bar{q}}}^{\dagger}
\end{array}
\right) , \lb{eq:NO1}
\ee
where
\be
O^{-1}(\theta_{\bo{k\bar{q}}})=\left(
\begin{array}{cc}
\cosh(\theta_{\bo{k\bar{q}}}) & -\sinh(\theta_{\bo{k\bar{q}}}) \\
-\sinh(\theta_{\bo{k\bar{q}}}) & \cosh(\theta_{\bo{k\bar{q}}})
\end{array}
\right). 
\ee

Now, we look for the parameter values of $\theta_{\bo{k\bar{q}}}$
for which the above $O^{-1}(\theta_{\bo{k\bar{q}}})$ reproduces
the matrix $O_{\bo{k\bar{q}}}^{-1}\hat{\bo{c}}_{\bo{k\bar{q}}}$ 
in Eq.~(\ref{eq:transR}) for real eigenvalues or in
Eq.~(\ref{eq:transC}) for complex ones. Simple calculations show 
\be
\theta_{\bo{k\bar{q}}}^{\rm I}  & = & 0 \, , \\
\cosh \theta_{\bo{k\bar{q}}}^{\rm R} & = & u_{\bo{k\bar{q}}} \, , \\
\sinh \theta_{\bo{k\bar{q}}}^{\rm R} & = & v_{\bo{k\bar{q}}} 
\ee
for real eigenvalues, and
\be
&\theta_{\bo{k\bar{q}}}^{\rm I}  =  - \frac{\pi}{4} \, ,& \\
&\frac{1}{\sqrt{2}} \left( \cosh \theta_{\bo{k\bar{q}}}^{\rm R} - i \sinh
\theta_{\bo{k\bar{q}}}^{\rm R} \right)  =  u_{\bo{k\bar{q}}} &
\ee
for complex ones.

Let us focus on the complex eigenvalue sector in which we have
explicitly 
\be
\hat{A}_{\bo{k\bar{q}_{\rm c}}} & = & \hat{V} \hat{c}_{\bo{k\bar{q}_{\rm
c}}}\hat{V}^{-1} \, , \\
\hat{B}_{\bo{k}-\bo{\bar{q}_{\rm c}}}^{\dag} & = & \hat{V}
\hat{c}_{\bo{k}-\bo{\bar{q}_{\rm c}}}^{\dag} \hat{V}^{-1} \, ,
\ee
and
\be
\hat{A}_{\bo{k\bar{q}_{\rm c}}}^{\dag} &=& V^{-1\dag}
\hat{c}_{\bo{k\bar{q}_{\rm c}}}^{\dag} V^{\dag} \nonumber \\
&=& - i \hat{A}_{\bo{k}-\bo{\bar{q}_{\rm c}}} \, ,  \lb{eq:NO2} \\
\hat{B}_{\bo{k}-\bo{\bar{q}_{\rm c}}} &=& V^{-1\dag}
\hat{c}_{\bo{k}-\bo{\bar{q}_{\rm c}}}V^{\dag} \nonumber \\ 
&=& - i \hat{B}_{\bo{k\bar{q}_{\rm c}}}^{\dag} \, . \lb{eq:NO3}
\ee

We first define new vacuum states by 
\be
|0\rangle_{A} &=& \hat{V}|0\rangle_{c}\, , \lb{zero1}\\
|0\rangle_{B} &=& \hat{V}^{-1\dag}|0\rangle_{c} \, , \lb{zero2}
\ee 
and
\be
{}_{A}\langle0| &=& {}_{c}\langle0|\hat{V}^{\dag}\, , \lb {zero3} \\
{}_{B}\langle0| &=& {}_{c}\langle0|\hat{V}^{-1}\, , \lb{zero4} 
\ee
where $|0\rangle_{c}$  is the vacuum of $\hat{c}_{\bo{k\bar{q}}}$. 
They are annihilated by $\hat{A}_{\bo{k\bar{q}_{\rm c}}}$,
 $\hat{A}_{\bo{k\bar{q}_{\rm c}}}^{\dag}$, $\hat{B}_{\bo{k\bar{q}_{\rm
 c}}}$ and $\hat{B}_{\bo{k\bar{q}_{\rm c}}}^{\dag}$ as follows: 
\be
\hat{A}_{\bo{k\bar{q}_{\rm c}}}|0\rangle_{A} =
\hat{A}_{\bo{k\bar{q}_{\rm c}}}^{\dag}|0\rangle_{A} =
0\, , \\
\hat{B}_{\bo{k\bar{q}_{\rm c}}}|0\rangle_{B} =
\hat{B}_{\bo{k\bar{q}_{\rm c}}}^{\dagger}|0\rangle_{B}= 0 \, ,
\ee
and 
\be
{}_{A}\langle0|\hat{A}_{\bo{k\bar{q}_{\rm c}}}=
{}_{A}\langle0|\hat{A}_{\bo{k\bar{q}_{\rm c}}}^{\dag} = 0 \, , \\
{}_{B}\langle0|\hat{B}_{\bo{k\bar{q}_{\rm c}}}=
{}_{B}\langle0|\hat{B}_{\bo{k\bar{q}_{\rm c}}}^{\dag} = 0 \, . 
\ee
From the commutation relations (\ref{eq;CCCR1})--(\ref{eq;CCCR3}), it
turns out that these states are eigenstates of the Hamiltonian
$\hat{H}_{0}$, 
\be
\hat{H}_{0}|0\rangle_{A} &=& - \left( \sum_{\bo{\bar{q}_{\rm c}}}
\frac{\hbar\omega_{\bo{k\bar{q}_{\rm c}}}^{(+)*}}{2} \right)
|0\rangle_{A} \, , \\
\hat{H}_{0}|0\rangle_{B} &=& - \left( \sum_{\bo{\bar{q}_{\rm c}}}
\frac{\hbar\omega_{\bo{k\bar{q}_{\rm c}}}^{(+)}}{2} \right) |0\rangle_{B}
\, .
\ee
Note the relations for ${}_{A}\langle0|$ and ${}_{B}\langle0|$\,,
\be
{}_{A}\langle0| \hat{H}_{0} &=& - \left( \sum_{\bo{\bar{q}_{\rm c}}}
\frac{\hbar\omega_{\bo{k\bar{q}_{\rm c}}}^{(+)}}{2} \right){}_{A}
\langle0| \, ,\\
{}_{B}\langle0|\hat{H}_{0} &=& - \left( \sum_{\bo{\bar{q}_{\rm c}}}
\frac{\hbar\omega_{\bo{k\bar{q}_{\rm c}}}^{(+)*}}{2} \right){}_{B}
\langle0|\, .
\ee

The commutation relations (\ref{eq;CCCR1}) suggest us to introduce
 the following excited states through cyclic operations of ${\hat
 B}^\dag$ on $|0\rangle_{A}$ and of ${\hat A}^\dag$ on $|0\rangle_{B}$:
\be
|N_{\rm C} \rangle_{A} &\equiv& |n_{\bo{\bar{q}_{{\rm c}\bo{1}}}}
 \cdots n_{\bo{\bar{q}_{\rm{c}\bo{i}}}} \rangle_{A} \nonumber \\
&=& \prod_{\bo{\bar{q}_{{\rm c}{\bo j}}}}
\sqrt{\frac{1}{n_{\bo{\bar{q}_{{\rm c}{\bo j}}}}!}}
(\hat{B}_{\bo{k\bar{q}_{{\rm c}{\bo j}}}}^{\dag})^{n_{\bo{\bar{q}_{{\rm
 c}{\bo j}}}}} |0\rangle_{A} \, , \label{eq:defNCA} \\
|N_{\rm C}\rangle_{B} &\equiv& |n_{\bo{\bar{q}_{\rm c1}}} \cdots
 n_{\bo{\bar{q}_{{\rm c}{\bo i}}}} \rangle_{B} \nonumber \\
&=& \prod_{\bo{\bar{q}_{{\rm c}{\bo j}}}}
 \sqrt{\frac{1}{n_{\bo{\bar{q}_{{\rm c}{\bo
 j}}}}!}}(\hat{A}_{\bo{k\bar{q}_{{\rm c}{\bo
 j}}}}^{\dag})^{n_{\bo{\bar{q}_{{\rm c}{\bo j}}}}} |0\rangle_{B} \, .
\label{eq:defNCB}
\ee
These excited states are also eigenstates of $\hat{H}_{0}$ as
\be
\hat{H}_{0}|N_{\rm C}\rangle_{A} 
&=& \sum_{\bo{\bar{q}_{\rm c}}} \left( n_{\bo{\bar{q}_{\rm c}}} \hbar
 \omega_{\bo{k\bar{q}_{\rm c}}}^{(+)} - \frac{\hbar
 \omega_{\bo{k\bar{q}_{\rm c}}}^{(+)*}}{2} \right) |N_{\rm C}
 \rangle_{A}\, , \nonumber\\ 
\\
\hat{H}_{0}|N_{\rm C} \rangle_{B} 
&=& \sum_{\bo{\bar{q}_{\rm c}}}\left(n_{\bo{\bar{q}_{\rm c}}} \hbar
\omega_{\bo{k\bar{q}_{\rm
c}}}^{(+)*}-\frac{\hbar\omega_{\bo{k\bar{q}_{\rm c}}}^{(+)}}{2} \right)
|N_{\rm C}\rangle_{B} \, .\nonumber \\
\ee

Let us evaluate ${}_{A}\langle0|0\rangle_{A}$, which is rewritten as
\be
{}_{A}\langle0|0\rangle_{A}&=&{}_{c}\langle0|\hat{W}|0\rangle_{c}
\nonumber \\ 
\hat{W}&=&\exp\Big{[} i \frac{\pi}{2} \sum_{\bo{\bar{q}_{\rm c}}} \left(
\hat{c}_{\bo{k\bar{q}_{\rm c}}} \hat{c}_{\bo{k}-\bo{\bar{q}_{\rm c}}} -
\hat{c}_{\bo{k\bar{q}_{\rm c}}}^{\dag} \hat{c}_{\bo{k}-\bo{\bar{q}_{\rm
c}}}^{\dag} \right) \Big{]} \, . \nonumber \\
\ee
Consider a state $|\xi\rangle$, 
\be
|\xi\rangle &\equiv& \exp\Big{[} \sum_{\bo{\bar{q}_{\rm c}}} - \xi
\left( \hat{c}_{\bo{k\bar{q}_{\rm c}}} \hat{c}_{\bo{k}-\bo{\bar{q}_{\rm
c}}} - \hat{c}_{\bo{k\bar{q}_{\rm c}}}^{\dag}
\hat{c}_{\bo{k}-\bo{\bar{q}_{\rm c}}}^{\dag} \right) \Big{]}
|0\rangle_{c} \nonumber \\ 
&=& \exp\Big{[} - \ln \cosh \xi + \sum_{\bo{\bar{q}_{\rm c}}}
\hat{c}_{\bo{k\bar{q}_{\rm c}}}^{\dag} \hat{c}_{\bo{k}-\bo{\bar{q}_{\rm
c}}}^{\dag} \tanh \xi \Big{]} |0\rangle_{c}  \, , \nonumber \\
&& \lb{eq:divergence}
\ee
then we have the relation ${}_{A}\langle0|0\rangle_{A}
={}_{c}\langle0|\xi=-i \pi/2 \rangle$. Obviously
${}_{c}\langle0|\xi\rangle$ diverges at the limit $\xi\to -i \pi/2$,
so does  ${}_{A}\langle0|0\rangle_{A}$. Similarly 
${}_{B}\langle0|0\rangle_{B}$ is also divergent.

Using Eqs.~(\ref{zero1})--(\ref{zero4}), one easily derives
\be
{}_{A}\langle0|0\rangle_{B}&=&1 \, , \\
{}_{B}\langle0|0\rangle_{A}&=&1 \, .
\ee 
The commutation relations in Eqs.~(\ref{eq;CCCR1})--(\ref{eq;CCCR3})
lead to
\be
& &{}_{A}\langle N_{\rm C}^{'}|N_{\rm C} \rangle_{B} =
\prod_{\bo{\bar{q}_{{\rm c}{\bo j}}}} 
\delta_{n'_{\bo{\bar{q}_{{\rm c}{\bo j}}}}n_{\bo{\bar{q}_{{\rm c}{\bo
j}}}}}\, , \\ 
& &{}_{B}\langle N_{\rm C}^{'}|N_{\rm C} \rangle_{A} =
\prod_{\bo{\bar{q}_{{\rm c}{\bo j}}}} 
\delta_{n'_{\bo{\bar{q}_{{\rm c}{\bo j}}}}n_{\bo{\bar{q}_{{\rm c}{\bo
j}}}}}\, . 
\ee

Let us rewrite the completeness relation in the complex mode sector,
using the complete set of $|N_{\rm C}\rangle_{c}$,
\be
{\bf 1}_{\rm C} = \sum_{N_{\rm C}} | N_{\rm C} \rangle_{c} \,\,{}_{c}
\langle N_{\rm C}|\, ,\lb{comset1}
\ee
where ${\bf 1}_{\rm C}$ is the identity operator in the complex sector
and 
\be
|N_{\rm C}\rangle_{c} \equiv \prod_{\bo{\bar{q}_{{\rm c}{\bo j}}}}
\sqrt{\frac{1}{n_{\bo{\bar{q}_{{\rm c}{\bo j}}}}!}}
(\hat{c}_{\bo{k\bar{q}_{{\rm c}{\bo j}}}}^{\dag})^{n_{\bo{\bar{q}_{{\rm
c}{\bo j}}}}}|0\rangle_{c}\, . 
\ee
Having $\hat{V}$ and $\hat{V}^{-1}$ operate on Eq.~(\ref{comset1}) from
the left and right, respectively, we obtain the completeness relation
using the states $| n_{\bo{\bar{q}_{{\rm c}{\bo 1}}}}, \cdots,
n_{\bo{\bar{q}_{{\rm c}{\bo i}}}} \rangle_{A}$ and ${}_{B}\langle
n_{\bo{\bar{q}_{{\rm c}{\bo 1}}}}, \cdots, n_{\bo{\bar{q}_{{\rm c}{\bo
i}}}}|$ as 
\be
{\bf 1}_{\rm C} &=& \sum_{N_{\rm C}} \hat{V} |N_{\rm C} \rangle_{c} \,\,
{}_{c}\langle N_{\rm C} |\hat{V}^{-1} \nonumber \\
&=& \sum_{N_{\rm C}}|N_{\rm C} \rangle_{A} \,\, {}_{B}\langle N_{\rm C}
|\, . 
\ee
Similarly another relation using the states $|N_{\rm C} \rangle_{B}$ and
${}_{A}\langle N_{\rm C}|$ follows:
\be
{\bf 1}_{\rm C} = \sum_{N_{\rm C}}| N_{\rm C} \rangle_{B} \,\, {}_{A}
\langle N_{\rm C}|.
\ee
One may say that a natural conjugate of $|N_{\rm C} \rangle_{A}$ is 
${}_{B}\langle N_{\rm C} |$ and vice versa \cite{MOSY}.

\section{Physical States}
In the previous section, we have ``diagonalized'' the unperturbed
Hamiltonian including complex eigenvalues and have found its
eigenstates. 
The state space is not a simple Fock one. We need to impose appropriate
conditions to construct a restricted physical state space. In QFT,
unstable behaviors of system are described in a stable picture such as
the Beliaev process.  We should now establish a stable particle picture
specified by the unperturbed Hamiltonian, and the decay processes are
described as the higher order of perturbation. We presume that unstable
behaviors of the BECs in optical lattices occur due to external
perturbation. 

As in Ref.~ \cite{MOSY}, we require the following physical state
conditions (PSCs).
\be
{\rm i})& &
\langle\overline{\Omega}|\hat{\Psi}(x)|\Omega\rangle=v(\bo{x}) \,,\nn\\ 
{\rm ii})&
&\langle\overline{\Omega}|\hat{\Psi}^{\dagger}(x)\hat{\Psi}(x)|\Omega
\rangle  \text{ is time-independent}\,, \nn\\ 
{\rm iii})& & \text{$\langle\overline{\Omega}|\hat{G}|\Omega\rangle$ is
real, when $\hat{G}$ is any Hermitian operator\,,}\nn\\ 
{\rm iv})& &\langle\overline{\Omega}|\Omega\rangle=1\nn\,,
\ee
where $\langle\overline{\Omega}|$ is the natural conjugate of
$|\Omega\rangle$. If $\langle\overline{\Omega}|$ and $|\Omega\rangle$
satisfy the above four conditions, we call them {\it physical
states}. The conditions {\rm i}) and {\rm ii}) mean that the order
parameter and density distribution are stationary without
perturbation. The condition {\rm iii}) guarantees that the expectation
value of any Hermitian operator can interpreted as physical
quantity. The condition {\rm iv}) is necessary for the probability
interpretation. The vacuum states which satisfy the PSCs are obtained 
as direct sum of $|0\rangle_{A}$ and $|0\rangle_{B}$\,,
\be
|0\rangle_{\oplus}
&\equiv& \frac{1}{\sqrt{2}}\left(|0\rangle_{A}\oplus|0\rangle_{B}\right)
\,,\\  
{}_{\oplus}\langle 0| 
&\equiv& \frac{1}{\sqrt{2}}\left(_{B}\langle 0|\oplus {}_{A}\langle
0|\right) . 
\ee 
The proof that these direct sum states satisfy PSCs is given in
Ref.~\cite{MOSY}. 
Here we add that the direct sum of the excited states
 $|N_{\rm C} \rangle_{A}$ and $|N_{\rm C} \rangle_{B}$
are also physical states,
\be
|N_{\rm C} \rangle_{\oplus}
&\equiv& \frac{1}{\sqrt{2}}\left(|N_{\rm C} \rangle_{A} \oplus
|N_{\rm C} \rangle_{B} \right) \, , \label{eq:ketNCOplus} \\
{}_{\oplus}\langle N_{\rm C}| 
&\equiv& \frac{1}{\sqrt{2}}\left(_{B}
\langle N_{\rm C}|\oplus {}_{A}\langle N_{\rm C} |\right)
\, . \label{eq:braNCOplus}
\ee

\section{Linear Response}

So far, we have developed the description of QFT with complex
eigenvalues. 
But complex eigenvalues are not directly connected with the
instability of a condensate. In this section, we discuss the dynamics of 
the system with complex eigenvalues, studying the response of a
condensate against external perturbation.  To derive theoretical
expressions is straightforward  in the linear
response theory (LRT) \cite{Kubo} with our formulation of QFT.
We also show numerical results of LRT and compare them with those from the
TDGP equation, concretely those from the discrete nonlinear Schr\"odinger
equation (DNSE) which is obtained by applying the tight-binding
approximation to the TDGP equation \cite{DNLS,DNLS2}.

\subsection{Formula}

The field operator $\hat{\Psi}$ is expanded in terms of the Wannier
functions as
$\hat{\Psi}=v_{\bo{k}}+\sum_{\bo{i}}\hat{a}_{\bo{ki}}w_{\bo{ki}}$. 
The particle number operator $\hat{N}=\int\! d^3x\, \hat{\Psi}^{\dag}
\hat{\Psi}$ is written as 
\be
\hat{N} &=& \sum_{\bo{i}} \rho_{\bo{i}}(t) \, , \\
\hat{\rho}_{\bo{i}}(t) &=& \hat{\rho}_{\bo{i}}^{(0)}(t) +
\hat{\rho}_{\bo{i}}^{(\mathrm{ex})}(t) \, , \\  
\hat{\rho}_{\bo{i}}^{(0)}(t) &=& n_{\rm c} + n_{\rm c}^{\frac{1}{2}}
e^{-i\bo{k\cdot x_{\bo{i}}}} \hat{a}_{\bo{ki}} (t) + n_{\rm
c}^{\frac{1}{2}} e^{i\bo{k\cdot x_{\bo{i}}}} \hat{a}_{\bo{ki}}^{\dag}
(t) \, , \nonumber \\
\\ 
\hat{\rho}_{\bo{i}}^{(\mathrm{ex})}(t) &=& \hat{a}_{\bo{ki}}^{\dag} (t) 
\hat{a}_{\bo{ki}} (t) \, , 
\ee
where $\rho_{\bo{i}}(t)$ is the particle density operator at the ${\bo
i}$-th site, and $\hat{a}_{\bo{ki}}(t)$ is 
\be
\hat{a}_{\bo{ki}}(t) &=& \sum_{\bo{\bar{q}_{\rm r}}}
\frac{e^{i(\bo{k}+\bo{\bar{q}_{\rm r}}) \cdot \bo{x}_{\bo{i}}}}{I_{\rm
s}^{\frac{1}{2}}} \hat{c}_{\bo{k\bar{q}_{\rm r}}} (t) +
\sum_{\bo{\bar{q}_{\rm c}}} \frac{e^{i(\bo{k}+\bo{\bar{q}_{\rm c}})
\cdot \bo{x}_{\bo{i}}}}{I_{\rm s}^{\frac{1}{2}}}
\hat{c}_{\bo{k\bar{q}_{\rm c}}} (t) \, , \nonumber \\
\ee
where
\be
\hat{c}_{\bo{k\bar{q}}_{\rm r}}(t) &=& u_{\bo{k\bar{q}_{\rm r}}}
\hat{b}_{\bo{k\bar{q}_{\rm r}}} e^{-i\omega_{\bo{k\bar{q}_{\rm
r}}}^{(+)} t} + v_{\bo{k\bar{q}_{\rm r}}}
\hat{b}_{\bo{k}-\bo{\bar{q}_{\rm r}}}^{\dag}
e^{i\omega_{\bo{k}-\bo{\bar{q}_{\rm r}}}^{(+)}t} \, , \nonumber \\
&& \\
\hat{c}_{\bo{k\bar{q}}_{\rm c}}(t) &=& u_{\bo{k\bar{q}_{\rm c}}}
\hat{A}_{\bo{k\bar{q}_{\rm c}}} e^{-i\omega_{\bo{k\bar{q}_{\rm
c}}}^{(+)}t} - iu_{\bo{k\bar{q}_{\rm c}}}^{*}
\hat{B}_{\bo{k}-\bo{\bar{q}_{\rm c}}}^{\dag}
e^{-i\omega_{\bo{k\bar{q}_{\rm c}}}^{(+)*}t} \, . \nonumber\\
&& 
\ee

We consider the external perturbation as  
\be
\hat{H}_{\mathrm{per}}(t) &=& \int \! d^3x \, \hat{\Psi}^{\dag}
 V_{\mathrm{per}}(\bo{x},t) \hat{\Psi} \nonumber \\
&=& \sum_{\bo{i,j}}\delta V_{\bo{ij}}(t)
\left(n_{\rm c}+n_{\rm c}^{\frac{1}{2}} e^{-i\bo{k\cdot x_{\bo{i}}}}
\hat{a}_{\bo{kj}} \right. \nonumber \\ 
& &\left.\hspace{0.8cm} {} + n_{\rm c}^{\frac{1}{2}} e^{i\bo{k\cdot
x}_{\bo{j}}} \hat{a}_{\bo{ki}}^{\dag} + \hat{a}_{\bo{ki}}^{\dag}
\hat{a}_{\bo{kj}} \right) \, , \lb{external perturbation} 
\ee
where
\be
\delta V_{\bo{ij}}(t) &=& \int\! d^3x\, w_{\bo{ki}}^{*}\delta
V_{\mathrm{per}}w_{\bo{kj}} \, . 
\ee
The function $\delta V_{\mathrm{per}}(\bo{x},t)$ represents
the time-dependent modification of trap. We use the on-site
approximation for $\delta V_{\bo{ij}}(t)$. The external perturbative
Hamiltonian (\ref{external perturbation}) becomes 
\be
\hat{H}_{\mathrm{per}}(t)=\sum_{\bo{i}}\delta
V_{\bo{i}}(t)\hat{\rho}_{\bo{i}}(t) \, , 
\ee
where we write $\delta V_{\bo{i}}(t)\equiv \delta V_{\bo{ii}}(t)$. 
 From the linear response theory (LRT) \cite{Kubo}, the change in the ${\bo
 i}$-th site particle density $\langle
 \delta\hat{\rho}_{\bo{i}}(t)\rangle$ is given as 
\be
\langle \delta \hat{\rho}_{\bo{i}} (t) \rangle &=& \langle \delta
\hat{\rho}_{\bo{i}}^{(0)} (t) \rangle + \langle \delta
\hat{\rho}_{\bo{i}}^{(\mathrm{ex})} (t) \rangle \, , \\
\langle \delta\hat{\rho}_{\bo{i}}^{(0)} (t) \rangle &=& \sum_{\bo{j}}
\frac{1}{i\hbar} \int_{-\infty}^{t} \! dt' \, \langle[
\hat{\rho}_{\bo{i}}^{(0)} (t), \hat{\rho}_{\bo{j}} (t') ] \rangle \delta
V_{\bo{j}} (t') \, , \lb{Cfunc} \nonumber \\
\\
\langle \delta \hat{\rho}_{\bo{i}}^{(\mathrm{ex})} (t) \rangle &=&
\sum_{\bo{j}} \frac{1}{i\hbar} \int_{-\infty}^{t}\! dt'\, \langle [
\hat{\rho}_{\bo{i}}^{(\mathrm{ex})} (t), \hat{\rho}_{\bo{j}} (t') ]
\rangle \delta V_{\bo{j}}(t') \, . \lb{Qfunc} \nonumber\\
\ee 
Here the expectation $ \langle \cdot \rangle$ is taken to be
 $\langle N|\cdot|N\rangle$ where $|N\rangle$ is a direct product of a
 Fock state for real mode  $|N_{\rm R}\rangle$ and
the physical state for complex mode $|N_{\rm C}\rangle_{\oplus}$
in Eq.~(\ref{eq:ketNCOplus}) with Eqs.~(\ref{eq:defNCA}) and
(\ref{eq:defNCB}): 
\be
|N\rangle &=& |N_{\rm R} \rangle |N_{\rm C} \rangle_{\oplus} \, ,
\lb{eq:NNRNC} 
\\
|N_{\rm R}\rangle &\equiv & |n_{\bo{\bar{q}_{{\rm r}{\bo 1}}}} \cdots
\rangle_{b} \nonumber \\ 
&=& \prod_{\bo{\bar{q}_{{\rm r}{\bo j}}}}
\sqrt{\frac{1}{n_{\bo{\bar{q}_{{\rm r}{\bo j}}}}!}}
(\hat{b}_{\bo{k\bar{q}_{{\rm r}{\bo j}}}}^{\dag})^{n_{\bo{\bar{q}_{{\rm
r}{\bo j}}}}} |0\rangle_{b} \, .
\ee 

The correlation function of Eq.~(\ref{Cfunc}) becomes
\be
& & \langle \big{[} \rho_{\bo{i}}^{(0)} (t), \rho_{\bo{j}}(t') \big{]}
\rangle \nonumber \\ 
&=& \frac{2in_{\rm c}}{I_{\rm s}} \sum_{\bar{q}}(u_{\bo{k\bar{q}}} +
v_{\bo{k\bar{q}}})^{2} \nonumber \\ 
& & \times \sin \left\{ \bo{\bar{q}} \cdot (\bo{x}_{\bo{i}} -
\bo{x}_{\bo{j}}) - \omega_{\bo{k\bar{q}}}^{(+)} (t-t') \right\}
\lb{response function 2}. 
\ee
Note that $\bar{\bo{q}}={\bo 0}$ mode is cancelled in Eq.~(\ref{response
function 2}) and does not show divergence in the limit of $\varepsilon
\to 0$. 

The correlation function of Eq.~(\ref{Qfunc}) becomes
\be
& & \langle \big{[} \rho_{\bo{i}}^{(\mathrm{ex})} (t), \rho_{\bo{j}}
(t') \big{]} \rangle \nonumber \\
&=& \frac{2i}{I_{\rm s}^{2}} \sum_{\bo{\bar{q}}_{1}, \bo{\bar{q}}_{2}}
\big\{ (n_{\bo{\bar{q}}_{1}} + 1) (n_{\bo{-\bar{q}}_{2}}+1) -
n_{\bo{\bar{q}}_{1}} n_{\bo{-\bar{q}}_{2}} \big\} \nonumber \\
& & \qquad \quad \times {\rm Re} \Big[ u_{{\bo k}\bar{\bo q}_{1}}
v_{{\bo k}\bar{\bo q}_{2}} (u_{{\bo k}{\bar{\bo q}_{1}}} v_{{\bo
k}{\bar{\bo q}}_{2}} + u_{{\bo k} \bar{\bo q}_{2}} v_{{\bo k}\bar{\bo
q}_{1}}) \nonumber \\ 
& & \qquad \quad \times \sin \Big{\{} ( \bo{\bar{q}}_{1} -
\bo{\bar{q}}_{2}) \cdot ( \bo{x}_{\bo{i}} - \bo{x}_{\bo{j}} ) \nonumber
\\ 
& & \qquad \qquad \quad - ( \omega_{\bo{k}\bo{\bar{q}}_{1}}^{(+)} +
\omega_{\bo{k}-\bo{\bar{q}}_{2}}^{(+)}) (t-t') \Big{\}} \Big]
\, . \label{eq:rhoex} 
\ee
This term gives rise to the singularity at $\bo{\bar{q}}={\bo 0}$ in the
limit $\varepsilon \to 0$. The infrared divergence caused by the
zero-mode singularity can be removed in the careful treatments of
renormalization \cite{Okumura1} or of the quantum coordinates
\cite{Okumura2}. 
In the numerical calculations below we drop the divergent term in 
Eq.~(\ref{eq:rhoex}) for simplicity, since the zero-energy contributions
are numerically  small after the treatments in
Refs.~\cite{Okumura1,Okumura2}.

\subsection{Numerical Result}
In this subsection, we show some numerical results of LRT with complex
 eigenvalues and compare those obtained from DNSE. 
Here we assume a system of one dimension in space for simplicity.

DNSE is given as follows
\be
i \hbar \frac{\partial}{\partial t} \Psi_{i} = -J (\Psi_{i-1} +
\Psi_{i+1}) + \left(\delta V_{i} +U|\Psi_{i}|^{2} \right) \Psi_{i}
\nonumber \,. \\ 
\ee
Here the quantity $|\Psi_{i}|^{2}$ represents the density of condensate
particle at the $i$-th site. 
Without external potential ($\delta V_{i}(t)=0$), DNSE has a stationary
solution
\be
\Psi_{i}^{(\mathrm{in})}(t)=n_{\rm c}^{\frac{1}{2}} e^{ikx_{i}}e^{- 
\frac{i}{\hbar} \mu t}\,. \lb{solutionDNSE}
\ee
We adopt this solution for the initial state of the wavefunction, and
define the density response as
\be
\delta n_{i} (t) = |\Psi_{i}(t)|^{2} - |\Psi_{i}^{(\mathrm{in})}(t)|^{2}
\, .
\ee
We focus on the external potential of the form
\be
\delta V_{i} (t) = S \exp \left( \frac{x_{i} - I_{\rm s}d/2} {\sigma}
\right)^{2} \theta (t-t_{0}) \, ,
\ee 
which pushes up the center of the lattice, switched on at $t=t_{0}$. 
The detailed form of the density response is given in
Appendix~\ref{appendix:response}. 
We have calculated the condensate particle density $n(x)$ numerically,
and have confirmed that $n(x)$ is symmetric under the conversion of
$x\to-x$, i.e., all the GP solutions for the one-dimensional system we
found satisfy $n(x)=n(-x)$. When $n(x)$ has the reflection symmetry, the 
Wannier functions and $J$ become real.
So we can set $\Theta=0$ which is the phase of $J$.

We set the parameter $S/J=0.0001$\,, $U/J=0.01$\,, $\sigma/d=0.2$ and
the total number of lattice sites $I_{\rm s}=51$ with the condensate
particle number per site $n_{\rm c}=2$. 

The density response in Eq.~(\ref{eq:Bdelrho})--(\ref{eq:Bdelrhoex})
 depends on the choice of the state $| N\rangle $ in
 Eq.~(\ref{eq:NNRNC}).  As we are 
 interested mainly in complex modes in this paper, we take the vacuum for
 real modes in our numerical calculations:
\be
|N_{\rm R}\rangle = | 0 \rangle_b \, .
\ee

First, we take the vacuum state of complex modes,
\be
|N_{\rm C}\rangle_{\oplus}=|0\rangle_{\oplus} \, . 
\ee

\begin{figure}[h]
\begin{center}
\includegraphics[width=1.00\linewidth]{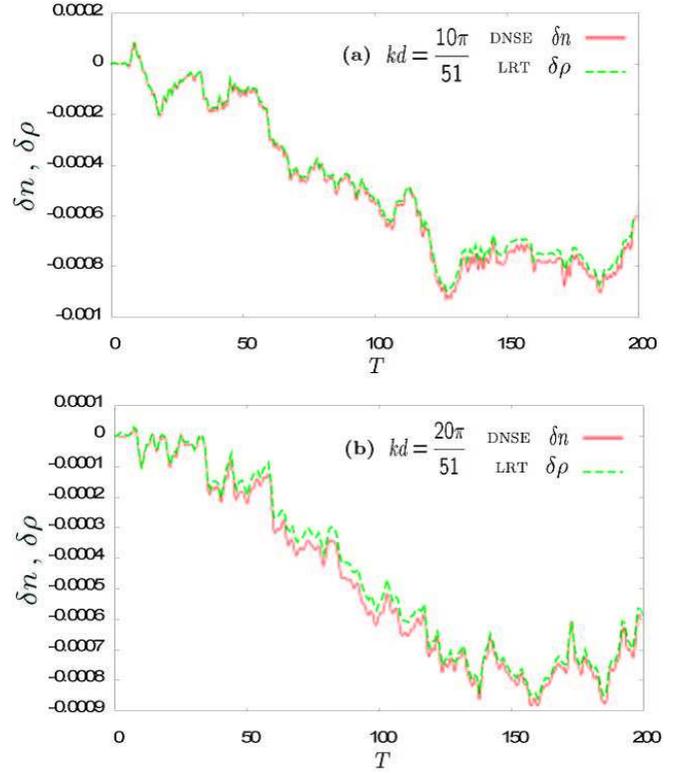}
\end{center}
\caption{\footnotesize{
(Color online)
Time evolution for the change of the density at the 10th site for (a)
 $kd=10\pi/51$ and (b) $kd=20\pi/51$ (region for real eigenvalues) with
$T=\frac{Jt}{\hbar}$.  
The solid line represents the result of DNSE $\delta n_{10}(T)$
 while the dashed line represents that of LRT $\langle 
\delta\hat{\rho}_{10}(T)\rangle$. }}
\label{compare1}
\end{figure}

The critical value of $k$, giving a boundary between real and complex
modes, is determined from  $kd=\pi/2$.
In Fig.~\ref{compare1}, the time evolution of the changes of the density 
at the 10th site are plotted for (a) $kd=10\pi/51$ and (b)
$kd=20\pi/51$, in both of the cases all the eigenvalues are real. 
We can see that the results of LRT fit that of DNSE with high precision,
as is expected.

\begin{figure}[h]
\begin{center}
\includegraphics[width=1.00\linewidth]{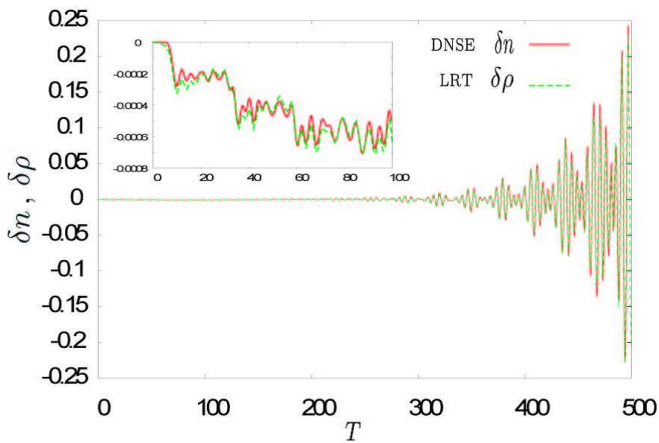}
\end{center}
\caption{\footnotesize{
(Color online)
Time evolution of the change of the density at the 10th site for
 $kd=26\pi/51$ (region for complex eigenvalues) with
 $T=Jt/\hbar$. 
The solid line represents the result of DNSE $\delta n_{10}(T)$
 while the dashed line represents that of LRT $\langle 
\delta\hat{\rho}_{10}(T)\rangle$.
}}
\label{compare2}
\end{figure}

Move to the case in which complex modes appear.  In Fig.~\ref{compare2},
the time evolution fo the change of the density at the 10th site is
plotted for $kd=26\pi/51$, this time the complex eigenvalues exist.  
One can find that the density response for small perturbation show the
characteristic behavior. 
The change of density for $kd=26\pi/51$ are larger than that of
$kd=10\pi/51$ or $kd=20\pi/51$ and grows exponentially.
This behavior is caused by the complex eigenvalues.
The result of LRT is in good agreement with that of DNSE again.
Recall that the operators $\hat{A}_{\bo{k\bar{q}_{\rm c}}}$,
$\hat{B}_{\bo{k\bar{q}_{\rm c}}}$ and the state $|0\rangle_{\oplus}$ are 
essential in our LRT formulation. 
The above agreement is not trivial at all when there are complex eigenmodes.

\begin{figure}[h]
\begin{center}
\includegraphics[width=1.00\linewidth]{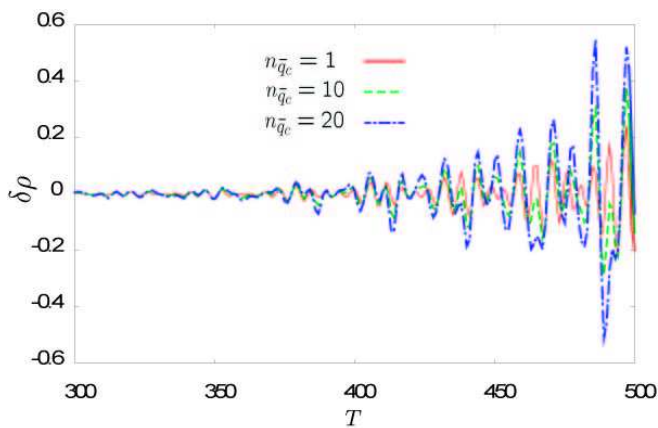}
\end{center}
\caption{\footnotesize{
(Color online)
Time evolution of the density at the 10th site $\langle
\delta\hat{\rho}_{10}(T) \rangle$ with $T=Jt/\hbar$, calculated with the
state defined in Eq.~(\ref{eq:nqc_state}) for the interval between $T=300$
and $T=500$. 
The solid, dashed and dashed-dotted lines correspond to $n_{\bar{q}_{\rm c}}=1,10$
and $20$, respectively. 
}}
\label{compare3}
\end{figure}

Next, we focus on one of the complex modes,
characterized by the Bloch vector $\bar{q}_{\rm c} = 2\pi/51$, and employ the
following singly excited state for $|N_{\rm C} \rangle_{\oplus}$
\be
|N_{\rm C} \rangle_{\oplus} &=& |0\cdots n_{\bar{q}_{\rm c}} \cdots 0
\rangle_{\oplus} \, .
\lb{eq:nqc_state}
\ee
In Fig.~\ref{compare3}, the quantity $\langle \delta\hat{\rho}_{10}(t)
\rangle$ is plotted as a function of time. 
As $n_{\bar{q}_{\rm c}}$ increases, the amplitude of the response
becomes larger and the exponentially diverging behavior becomes
prominent.  
Thus it seems that the excitations of the complex modes hasten the
collapse of condensates.

\section{Summary}

We have investigated the dynamical instability of Bose-Einstein
condensation flowing in an optical lattice.
The formulation in this paper is based on quantum field theory (QFT).
The field operator must include complex modes for the dynamically
unstable system, otherwise the canonical commutation relations would be
violated. 

We have attempted to diagonalize the unperturbed Hamiltonian
 under the tight-binding approximation, but it is not diagonalizable
in the conventional bosonic representation. Nevertheless one can find 
its eigenstates, i.e., the vacuum and excited states
in the complex mode sectors.
Then, the physical state conditions (PSCs) were introduced 
to restrict the state space, so that we can 
 start with the consistent stable particle in QFT.

As an application of our formulation to the problem of the dynamical
 instability, we consider the response of a condensate
against external perturbation in the regime of the linear response
theory (LRT). The numerical results of LRT with complex modes
are to be compared with those from the discrete Schr\"odinger equation
 (DNSE). 
It is remarkable that both of the results coincide with each other
although the two methods are quite different.
The state $|0\rangle_{\oplus}$ and the operators 
$\hat{A}_{\bo{k\bar{q}_{\rm c}}}$ and  $\hat{B}_{\bo{k\bar{q}_{\rm c}}}$ 
are crucial for our formulation of LRT.  
It is an interesting observation that the excited state in the complex
mode sector hastens the collapse of the condensate in comparison with
the vacuum state.

\begin{acknowledgments}
M.M. is supported partially by the Grant-in-Aid for The 21st
Century COE Program (Physics of Self-organization Systems) at Waseda
University.
This work is partly supported by a Grant-in-Aid for Scientific Research
(C) (No.~17540364) from the Japan Society for the Promotion of Science,
for Young Scientists (B) (No.~17740258) and for
Priority Area Research (B) (No.~13135221) both from the Ministry of
Education, Culture, Sports, Science and Technology, Japan.
\end{acknowledgments}

\appendix

\section{Eigenfunctions of the Bogoliubov--de Gennes equation}

In this Appendix, we rephrase the contents in Sec.~\ref{sec:Hdiag}
from the viewpoint of the Bogoliubov-de Gennes (BdG) equation. 
The QFT formalism on the trapped BECs with vortices in Ref.~\cite{MOSY}
is based on the BdG equation.

The relevant BdG equation is, in the doublet notation,
\be
T x_{n}(\bo{x})=\hbar\omega_{n}x_{n}(\bo{x})
\lb{eq:BdGeq}
\ee
where
\be
T&=&\left(
\begin{array}{cc}
L & M \\
-M^{*} & -L
\end{array}
\right) \, , \\
x_{n}(\bo{x})&=&\left( 
\begin{array}{c}
f_{n}(\bo{x}) \\
g_{n}(\bo{x})
\end{array}
\right)  \, ,\\
L&=&K+V_{\mathrm{opt}}-\mu+2g|v(\bo{x})|^{2} \, ,\\
M&=&gv^{2}(\bo{x}) \, .
\ee

The operator $T$ has the pseudo-Hermitian property of
\be
\sigma_3 T^\dagger \sigma_3 = T \, 
\ee
correspondingly to Eq.~(\ref{eq:pseudT}).  This leads us to
define the following inner product for an arbitrary pair of doublets,
\be
(r,s)&\equiv& \int d^{3}x\, r^{\dagger}(\bo{x})\sigma_{3}s(\bo{x}) \\
&=&\int \! d^{3}x \big{[}r_{1}^{*}(\bo{x}) s_{1}(\bo{x}) -
r_{2}^{*}(\bo{x}) s_{2} (\bo{x}) \big{]}, 
\ee
where
\be
r(\bo{x}) = \left( 
\begin{array}{c}
r_{1}(\bo{x}) \\
r_{2}(\bo{x})
\end{array}
\right) \, , \qquad 
s(\bo{x}) = \left( 
\begin{array}{c}
s_{1}(\bo{x}) \\
s_{2}(\bo{x})
\end{array}
\right)\, .
\ee
We may also define a (squared) ``norm'' of $r$ as $||r||^{2}\equiv
(r,r)$, which is  not positive-definite. One easily obtains
\be
(\hbar\omega_{n}-\hbar\omega_{n'}^{*})(x_{n'},x_{n})=0.\lb{eq:othrel}
\ee

As a counterpart of Eq.~(\ref{eq:sigma1T}) we find the relation,
\be
\sigma_{1}T^{*}\sigma_{1}=-T \, .
\ee
It turns out that
for any eigenvector $x_{n}(\bo{x})$ whose eigenvalue is denoted by
$\hbar\omega_{n}$
the doublet $y_{n}(\bo{x})=\sigma_{1}x_{n}^{*}(\bo{x})$ 
becomes an eigenvector with the eigenvalue $-\hbar\omega_{n}^{*}$.

When the eigenvalues are real, we have the following orthonormal
relations, consistent with Eq.~(\ref{eq:othrel}):
\be
(x_{n'},x_{n})&=&\delta_{n'n} \lb{eq:norm1},\\
(y_{n'},y_{n})&=&-\delta_{n'n} ,\lb{eq:norm2}\\
(y_{n'},x_{n})&=&0.\lb{eq:othogonal}
\ee

Complex modes appear in a pair for the BdG equation (\ref{eq:BdGeq}),
i.e., any eigenstate $\zeta_{m}$ belonging to a complex eigenvalue
$\hbar\omega_{m}$ is accompanied by another eigenstate $\eta_\ell$ whose 
eigenvalue $\hbar \omega_\ell$ is a complex conjugate of
$\hbar\omega_{m}$, $\hbar\omega_{\ell}=\hbar\omega_{m}^{*}$. This fact
is shown in constructing eigenstates explicitly in
Sec.~\ref{sec:Hdiag}. 
The  ``norm'' of the eigenstates of complex eigenvalues 
is zero,
\be
\|\zeta_{m}\|^{2}= \|\eta_{\ell}\|^2=0 \lb{eq:zeronorm} \, ,
\ee 
since  $(\hbar\omega_{m}-\hbar\omega_{m}^{*})$ is not zero in
Eq.~(\ref{eq:othrel}). The ``zero norm'' is a necessary condition for
the emergence of complex eigenvalues. 
The pair of the eigenvectors $\zeta_{m}$ and $\eta_{\ell}$ are not
orthogonal to each other in general,  
\be
(\eta_{\ell},\zeta_{m}) \neq 0 \lb{eq:nonothogonal}
\ee
as there is a vanishing factor of
$(\hbar\omega_{m}-\hbar\omega_{\ell}^{*})$ on the left-hand side of
Eq.~(\ref{eq:othrel}). 

Let us expand the field operator in the doublet notation in terms of the
eigenfunctions of the BdG equation,
\be
\hat{\Phi}(\bo{x},t) &=& \sum_{\bo{\bar{q}_{\rm r}}} \left[
x_{\bo{k\bar{q}_{\rm r}}} (\bo{x}) \hat{b}_{\bo{k\bar{q}_{\rm r}}} (t)
+ y_{\bo{k\bar{q}_{\rm r}}}(\bo{x}) \hat{b}_{\bo{k\bar{q}_{\rm
r}}}^{\dag} (t) \right]
\nonumber \\
&& {} + \sum_{\bo{\bar{q}_{\rm c}}} \left[ \zeta_{\bo{k\bar{q}_{\rm r}}}
(\bo{x}) \hat{A}_{\bo{k\bar{q}_{\rm r}}} (t) + \eta_{\bo{k\bar{q}_{\rm
r}}} (\bo{x}) \hat{B}_{\bo{k\bar{q}_{\rm r}}}^{\dag} (t) \right] \,,
\nonumber\\ 
\ee
where
\be
\hat{\Phi}(\bo{x},t)=\left( 
\begin{array}{c}
\hat{\phi}(\bo{x},t)\\
\hat{\phi}^{\dag}(\bo{x},t)
\end{array}
\right)  \, ,
\ee
as in the case of a BEC with a vortex \cite{MOSY}.
Comparing this expansion with Eq.~(\ref{eq:fieldexpre}) and its Hermitian
conjugate, we have
\be
x_{\bo{k\bar{q}_{\rm r}}}(\bo{x}) &=& \left( 
\begin{array}{c}
u_{\bo{k\bar{q}_{\rm r}}}f_{\bo{k\bar{q}_{\rm r}}}(\bo{x})\\
v_{\bo{k\bar{q}_{\rm r}}}f_{\bo{k-\bar{q}_{\rm r}}}^{*}(\bo{x})
\end{array}
\right)\,,  \\
y_{\bo{k\bar{q}_{\rm r}}} (\bo{x}) &=& \left( 
\begin{array}{c}
v_{\bo{k\bar{q}_{\rm r}}}f_{\bo{k\bar{q}_{\rm r}}}(\bo{x}) \\
u_{\bo{k\bar{q}_{\rm r}}}f_{\bo{k}-\bo{\bar{q}_{\rm r}}}^{*}(\bo{x})
\end{array}
\right) \,,\\
\zeta_{\bo{k\bar{q}_{\rm c}}}(\bo{x}) &=& \left( 
\begin{array}{c}
u_{\bo{k\bar{q}_{\rm c}}} f_{\bo{k\bar{q}_{\rm c}}}(\bo{x})\\
-iu^{*}_{\bo{k-\bar{q}_{\rm c}}}f_{\bo{k}-\bo{\bar{q}_{\rm
 c}}}^{*} (\bo{x}) 
\end{array}
\right) \,,\\
 \eta_{\bo{k\bar{q}_{\rm c}}}(\bo{x}) &=& \left( 
\begin{array}{c}
-iu^{*}_{\bo{k-\bar{q}_{\rm c}}}f_{\bo{k-\bar{q}_{\rm c}}}(\bo{x})\\
u_{\bo{k\bar{q}_{\rm c}}}f_{\bo{k}\bo{\bar{q}_{\rm c}}}^{*}(\bo{x})
\end{array}
\right).
\ee
It is straightforward to check the following orthonormal relations,
\be
(x_{\bo{k\bar{q}_{\rm r}'}},x_{\bo{k\bar{q}_{\rm r}}})
&=& \delta_{\bo{\bar{q}_{\rm r}\bar{q}_{\rm r}'}} \, ,\\
(y_{\bo{k\bar{q}_{\rm r}'}},y_{\bo{k\bar{q}_{\rm r}}})
&=& - \delta_{\bo{\bar{q}_{\rm r}\bar{q}_{\rm r}'}} \, , \\
(y_{\bo{k\bar{q}_{\rm r}'}},x_{\bo{k\bar{q}_{\rm r}}}) &=& 0 \, ,
\ee
and
\be
(\zeta_{\bo{k\bar{q}_{\rm c}'}},\zeta_{\bo{k\bar{q}_{\rm c}}}) &=& 0 \,
,\\ 
(\eta_{\bo{k\bar{q}_{\rm c}'}}, \eta_{\bo{k\bar{q}_{\rm c}}}) &=& 0 \, ,
\\ 
(\eta_{\bo{k\bar{q}_{\rm c}'}}, \zeta_{\bo{k\bar{q}_{\rm c}}})
&=& i \delta_{\bo{\bar{q}_{\rm c}\bar{q}_{\rm c}'}} \,
. \lb{eq:wnonothogonal}
\ee

\section{Expression of Density Response} \lb{appendix:response}
In this Appendix, we give the detailed expression of the density response
$\langle \delta\hat{\rho}_{i}(t)\rangle$. The state by which the
expectation $\langle\cdot \rangle$ is taken is found in
Eq.~(\ref{eq:NNRNC}). 
We restrict ourselves to the case of one dimension in space.

The external perturbation is given as
\be
\delta V_{i} (t) = S \exp \left( \frac{x_{i}-I_{\rm s}d/2} {\sigma}
\right)^{2} \theta(t-t_{0}) \, .
\ee
Then the expression of the density response becomes
\be
\langle \delta\hat{\rho}_{i} (t) \rangle = \langle
\delta\hat{\rho}_{i}^{(0)} (t) \rangle + \langle \delta
\hat{\rho}_{i}^{\rm (ex)}(t) \rangle \, , \label{eq:Bdelrho}
\ee
where
\begin{widetext}
\be
\langle \delta \hat{\rho}_{i}^{(0)} (t) \rangle &=& \frac{2S}{\hbar
 I_{\rm s}} \sum_{j,\bar{q}} \exp \left( \frac{x_{j}-I_{\rm s}d/2}
{\sigma}\right)^{2} \frac{n_{\rm
 c}(u_{k\bar{q}}+v_{k\bar{q}})^{2}}{\omega_{k\bar{q}}^{(+)}}
 \Big[ \cos \left\{ \bar{q}(x_{i}-x_{j}) - \omega_{k\bar{q}}^{(+)}
 (t-t_{0}) \right\} - \cos\{ \bar{q}(x_{i}-x_{j}) \} \Big] \, ,\nonumber
 \\ 
\label{eq:Bdelrho0}\\
\langle \delta \hat{\rho}_{i}^{(\mathrm{ex})} (t) \rangle 
&=& \frac{2S}{\hbar I_{\rm s}} \sum_{j} \sum_{\bar{q}_{1},\bar{q}_{2}
 \neq 0} \exp \left( \frac{x_{j}-I_{s}d/2}{\sigma} \right)^{2} \big\{
 (n_{{\bar{q}}_{1}}+1) (n_{{-\bar{q}}_{2}}+1) - n_{{\bar{q}}_{1}}
 n_{{-\bar{q}}_{2}} \big\} \nonumber \\
& & \times {\rm Re} \bigg{[} \frac{u_{k\bar{q}_{1}} v_{k\bar{q}_{2}}
 (u_{k\bar{q}_{1}} v_{k\bar{q}_{2}} + u_{kq_{2}}
 v_{kq_{1}})}{(\omega_{k\bar{q}_{1}}^{(+)} +
 \omega_{k-\bar{q}_{2}}^{(+)}) I_{s} } \Big{\{} \cos \big{(} (
 \bar{q}_{1} - \bar{q}_{2} ) ( x_{i}-x_{j} ) -
 (\omega_{k\bar{q}_{1}}^{(+)} + \omega_{k-\bar{q}_{2}}^{(+)}) (t-t_{0})
 \big{)} \nonumber \\
& &\hspace{6cm} {} - \cos \big( (\bar{q}_{1} - \bar{q}_{2} )(x_{i} -
 x_{j}) \big) \Big{\}} \bigg{]} \, . 
\label{eq:Bdelrhoex}
\ee 
\end{widetext}

\newpage 

\end{document}